\documentclass[a4paper,10pt]{article}
\pdfoutput=1
\usepackage[english]{babel}
\usepackage[utf8]{inputenc}
\usepackage[nottoc]{tocbibind}
\usepackage[affil-it]{authblk}
\usepackage{amsmath}%
\usepackage{amsfonts}%
\usepackage{amssymb}%
\usepackage[a4paper]{geometry}
\usepackage{subcaption}
\usepackage{graphicx}
\usepackage{mathrsfs}
\usepackage{bm}
\usepackage{bbold}
\usepackage{slashed}
\usepackage[hidelinks]{hyperref}
\usepackage{mathtools}
\usepackage{csquotes}
\usepackage[title]{appendix}
\hypersetup{
   colorlinks = true,
    linkcolor = {black},
    anchorcolor = {green},
    citecolor = {red},
    filecolor = {cyan},
    menucolor = {red},
    runcolor = {cyan},
    urlcolor = {magenta}
}

\newcommand{\R}{{\mathbb{R}}}

\newcommand{\Z}{{\mathbb{Z}}}
\newcommand{\N}{{\mathbb{N}}}

\newcommand{\beq}{\begin{equation}}
\newcommand{\eeq}{\end{equation}}
\newcommand{\bea}{\begin{eqnarray}}
\newcommand{\eea}{\end{eqnarray}}
\newcommand{\bal}{\begin{align}}
\newcommand{\eal}{\end{align}}

\newcommand{\bdy}{\partial}

\newcommand{\wt}{\widetilde}

\newcommand{\su}{{\mathfrak{su}}}

\newcommand{\tr}{{\rm tr}\, }

\newcommand{\bigslant}[2]{{\raisebox{.2em}{$#1$}\left/\raisebox{-.2em}{$#2$}\right.}}

\renewcommand{\epsilon}{\varepsilon}

\newtheorem{theorem}{Theorem}

\newtheorem{lemma}[theorem]{Lemma}

\makeatletter
\newcommand{\Spvek}[2][r]{%
  \gdef\@VORNE{1}
  \left(\hskip-\arraycolsep%
    \begin{array}{#1}\vekSp@lten{#2}\end{array}%
  \hskip-\arraycolsep\right)}

\def\vekSp@lten#1{\xvekSp@lten#1;vekL@stLine;}
\def\vekL@stLine{vekL@stLine}
\def\xvekSp@lten#1;{\def\temp{#1}%
  \ifx\temp\vekL@stLine
  \else
    \ifnum\@VORNE=1\gdef\@VORNE{0}
    \else\@arraycr\fi%
    #1%
    \expandafter\xvekSp@lten
  \fi}
\makeatother

\title{Skyrmions from calorons}
\author{Josh Cork\footnote{email address: jscork@hotmail.co.uk}}
\affil{School of Mathematics, University of Leeds,\\
Woodhouse Lane, Leeds, LS2 9JT, U.K.}
\date{\today}
\begin{document}
\maketitle
\begin{abstract}
We derive a one-parameter family of gauged Skyrme models from Yang-Mills theory on $S^1\times\R^3$, in which skyrmions are well-approximated by calorons and monopoles. In particular we study the spherically symmetric solutions to the model with two distinct classes of boundary conditions, and compare them to calorons and monopoles. Calorons interpolate between instantons and monopoles in certain limits, and we observe similar behaviour in the constructed gauged Skyrme model in the weak and strong coupling limits. This comparison of calorons, monopoles, and skyrmions may be a way to further understand the apparent relationships between skyrmions and monopoles on $\R^3$.
\end{abstract}
\hypersetup{
    linkcolor = {blue}
}
\section{Introduction}
There has long been interest in the topological soliton solutions \cite{MantonSutcliffe2004,shnir2018topsolitons} of the Skyrme model \cite{skyrme1962nucl}, also known as \textit{skyrmions}. In its simplest form, the Skyrme model is very mathematically neat, with interesting geometric properties \cite{Manton1987geometry}, however its main interest lies in the physical interpretation as a non-linear field theory of pions with the topological charge of a skyrmion, an integer, identified as the number of baryons. Despite its simplicity, the model has a few crucial drawbacks. For one, the Skyrme model is not a BPS theory, in as much as its (non-trivial) critical points cannot attain the topological energy bound. This leads to skyrmions having rather large binding energies, which is in stark disagreement with physical expectations. Additionally, explicit solutions to the Skyrme field equation are very hard to come by. Although some analytic solutions are known on compact manifolds (see for example \cite{AlvarezCanforaDimakisPaliathanasis2017integrability}), for the non-compact $\R^3$, there is, for example, still no known explicit solution describing a skyrmion of charge $1$, despite the fact it is known to exist \cite{esteban1986SkyrmeVarSol,esteban2004existence}. Even less is known about existence beyond charge $1$, but there has nevertheless been significant work in constructing numerical solutions \cite{battyesutcliffe1997symmetricskyme,battye2001solitonic,battye2002skyrmions,BraatenTownsendCarson1990,FeistLauManton2013skyrmions}.

In 1989, Atiyah and Manton \cite{AtiyahManton1989} proposed a simple method for approximating skyrmions: from an $SU(2)$ instanton on $\R^4$ one may calculate its holonomy along all lines in one direction in $\R^4$, yielding a function $U:\R^3\longrightarrow SU(2)$ satisfying all of the correct boundary conditions to qualify as a Skyrme field. This approach has been demonstrated to have excellent accuracy and success, for example, when compared to the numerical solutions, the energies of the instanton generated Skyrme fields typically lie within $1\%$ of the `true energy' given by the numerical solutions, moreover, many of the symmetric examples of skyrmions that have been found numerically have been reproduced and confirmed by this instanton construction \cite{AtiyahManton1989,LeeseManton1994stable,MantonSutcliffe1995skyrme,SingerSutcliffe1999,sutcliffe2004Buckyball}.

For quite some time, nobody had been able to provide a good explanation as to why this construction appeared to approximate skyrmions so well. It wasn't until 2010 that Sutcliffe supplied the answer \cite{sutcliffe2010skyrmions}. Drawing inspiration from a Skyrme model introduced by Sakai and Sugimoto \cite{SakaiSugimoto2005} in the context of holographic QCD, Sutcliffe's approach was to employ a mode expansion of the instanton gauge field, allowing one to rewrite the Yang-Mills lagrangian as the energy for a static Skyrme model, where the Skyrme field is coupled to an infinite tower of vector mesons. The ordinary Skyrme model is a truncation of this model in which all of the vector mesons are artificially set to zero. Sutcliffe's construction effectively killed three birds with one stone: not only did it explain the success of the instanton approximation, but it has the fortunate advantage that the fully extended Skyrme model with all vector mesons is a BPS theory, for which the instanton Skyrme fields are exact solutions. Furthermore, even the inclusion of a finite number of vector mesons significantly reduces the binding energies, leading to more realistic theories \cite{sutcliffe2011skyrmionsTruncated}. The unfortunate pay-off is that their inclusion is a numerically taxing problem, although some important progress has been made in this area \cite{NayaSutcliffe2018skyrmions}.

In recent years, there has been a resurgence of interest in \textit{calorons} -- instantons on $S^1\times\R^3$ -- for a variety of reasons \cite{CharbonneauHurtubise2008Rat.map,GarlandMurray1988,HekmatiMurrayVozzo2012CalCor,KraanVanBaal1998Tdual,kraanVanBaal1998monopoleconsts,KraanVanBaal1998,norbury2000periodic}. Among the interesting aspects of calorons is their relationship to monopoles on $\R^3$ \cite{GarlandMurray1988,kraanVanBaal1998monopoleconsts}, and how they exhibit an interpolation between instantons on $\R^4$ and monopoles on $\R^3$ \cite{Harland2007,LeeLu1998,Rossi1979,Ward2004}. Furthermore, they have a notable physical relevance, having been linked to the problem of quark confinement \cite{AlkoferGreensite2007quark,vanBaal2007review}. Replacing the instanton in the Atiyah-Manton ansatz with a caloron, and taking the holonomy along a line in $\R^3$, yields a periodic skyrmion, or \textit{Skyrme chain} \cite{HarlandWard2008chains}. An alternative approach is to instead calculate caloron holonomies in the $S^1$ direction, resulting in an ordinary Skyrme field on $\R^3$. This was investigated very soon after Atiyah and Manton's original paper in \cite{EskolaKajantie1989skycal,NowakZahed1989skycal} (and long before the Sutcliffe construction). This again showed to be a good approximation of skyrmions. However, there is a problem with using calorons to approximate skyrmions in this way. Apart from the spherically symmetric cases, performing a gauge transformation of the caloron is not a symmetry of the resulting Skyrme field. In fact, this \textit{gauge variance} is a general feature of trying to construct skyrmions from instanton holonomies around circles (for example as in \cite{MantonSamols1990skyrmions}).

This paper is a reassessment of the work of \cite{EskolaKajantie1989skycal,NowakZahed1989skycal}, in light of Sutcliffe's model. As we shall see, this problem of missing gauge invariance is no longer apparent when we incorporate an $SU(2)$ gauge field into the model. This means that calorons should really be compared to \textit{gauged skyrmions}. Some work regarding gauged Skyrme models has been done before, for example in \cite{ArthurTchrakian1996GaugedSky,BrihayeHartmannTchhrakian2001MonopolesGaugeSky,BrihayeHillZachos2004BoundingGaugeSkyMass,PietteTchrakian2000static}. Another advantage of looking at calorons in this way is down to their relationship to monopoles. Indeed, calorons provide a natural way to compare monopoles with skyrmions, and this consideration, upon further investigation, may be a way to explain the curious similarities between monopoles and skyrmions \cite{HoughtonMantonSutcliffe1998rational}. Some attempts to do this were made in \cite{GrigorievSutcliffeTchrakian2002skyrmedmon,PaturyanTchrakian2004monopoleSky} by adding a `Skyrme-term' to the Yang-Mills-Higgs energy. In contrast, our approach does not make any modifications to the underlying theories, and therefore more direct comparisons can be made.

A brief outline of this paper is as follows. In section \ref{sec.Skyrme-mod}, we review the ordinary Skyrme model, and the Atiyah-Manton-Sutcliffe construction from instantons. In section \ref{sec.Sky-from-cal} we derive a one-parameter-family of gauged Skyrme models from a compactified Yang-Mills theory, and study its large parameter limit and its topological bounds. Section \ref{sec.numerics} is dedicated to studying the spherically symmetric critical points of the gauged Skyrme models. We study two types of boundary conditions -- `monopole-like' and `instanton-like' -- specifically for the purpose of making a comparison with calorons which possess the same boundary conditions. It is observed that the gauged skyrmions exhibit an interpolation between the monopole-like and instanton-like boundary conditions with respect to changing the parameter in the model. Finally, in section \ref{sec.gauged-sky-to-sky} we compare calorons and gauged skyrmions to ordinary skyrmions in the spherically symmetric case.
\section{Skyrme models}\label{sec.Skyrme-mod}
An $SU(2)$ Skyrme model is a $3$-dimensional field theory, whose field content is a function $U:\R^3\longrightarrow SU(2)$ called the \textbf{Skyrme field}. The simplest form of a static energy for the Skyrme model is given by
\begin{align}\label{Ordinary_Skyrme_energy}
    E_S=\int\left(c_0|U^{-1}dU|^2+c_1|U^{-1}dU\wedge U^{-1}dU|^2\right)d^3x,
\end{align}
where here $|\cdot|$ is the norm of the inner product on $\Omega(\R^3,\su(2))$: $\langle\zeta,\eta\rangle d^3x=-\tr(\zeta\wedge\star_3\eta)$, with $\star_3$ denoting the Hodge-star on $\R^3$, and $c_1,c_2\in\R^+$ are arbitrary constants representative of a choice of length and energy units. The $SU(2)$ Skyrme model is a model of nuclear physics, and in particular, is a theory of pions. The pion fields are manifested as a vector valued field $\vec{\pi}:\R^3\longrightarrow\R^3$ seen in the Skyrme field's $SU(2)$ expansion:
\begin{align}
    U(\vec{x})=\phi(\vec{x})\mathbb{1}+\imath\vec{\pi}(\vec{x})\cdot\vec{\sigma},
\end{align}
where $\vec{\sigma}=(\sigma^1,\sigma^2,\sigma^3)$ are the Pauli matrices. The function $\phi:\R^3\longrightarrow\R$ is a scalar field which has less physical significance than the pion fields. Critical points of the Skyrme energy functional (\ref{Ordinary_Skyrme_energy}) satisfy the \textit{Skyrme field equation}
\begin{align}\label{Skyrme-field-equation}
    \sum_{i,j}\bdy_i\left(c_0U^{-1}\bdy_iU+c_1\left[U^{-1}\bdy_jU,[U^{-1}\bdy_iU,U^{-1}\bdy_jU]\right]\right)=0.
\end{align}
The boundary condition $U\to\mathrm{const}$ as $|\vec{x}|\to\infty$ is usually imposed, and since $E_S$ is invariant under left multiplication of $U$ by constant $SU(2)$ matrices, the boundary condition $U\to\mathbb{1}$ as $|\vec{x}|\to\infty$ may be chosen without loss of generality. It is suspected that the condition of finite energy implies this boundary condition, but this is still yet to be proven. Critical points with this boundary condition are called \textbf{skyrmions}. The boundary condition implies that a skyrmion descends to a map $U':S^3\longrightarrow SU(2)$, which is characterised by an integer degree $\mathcal{B}\in\pi_1(SU(2))\cong\Z$, called the \textit{Skyrme charge}, and is physically interpreted as the \textbf{baryon number}. This has the integral formula
\begin{align}\label{baryon-number_integral}
    \mathcal{B}=\dfrac{1}{24\pi^2}\int\tr(U^{-1}dU\wedge U^{-1}dU\wedge U^{-1}dU).
\end{align}
The Skyrme energy (\ref{Ordinary_Skyrme_energy}) has a topological bound, originally due to Fadeev \cite{faddeev1976some}, given by
\begin{align}
    E_S\geq 48\pi^2\sqrt{c_0c_1}|\mathcal{B}|.\label{Fadeev-bound}
\end{align}
It is a straightforward exercise to show that no critical points of (\ref{Ordinary_Skyrme_energy}) besides constants ($\mathcal{B}=0$) can saturate this bound \cite{Manton1987geometry}.
\subsection{The Atiyah-Manton-Sutcliffe construction}
Let $D^A$ be a connection on an $SU(2)$ bundle over $\R^4$ with connection $1$-form $A\in\Omega^1(\R^4,\su(2))$. Giving $\R^4$ coordinates $(x^0,x^1,x^2,x^3)\equiv(x^0,\vec{x})$, its holonomy in the $x^0$-direction may be calculated via the solution $\Omega:\R^4\longrightarrow SU(2)$ to the parallel transport equation
\begin{align}
    \bdy_0\Omega+A_0\Omega=0,\quad\lim_{x^0\to-\infty}\Omega=\mathbb{1},
\end{align}
as $U(\vec{x})=\lim_{x^0\to\infty}\Omega(x^0,\vec{x})$. Formally this is given by the path-ordered exponential
\begin{align}
    U(\vec{x})=\mathcal{P}\exp\left(-\int_{-\infty}^\infty A_0(z,\vec{x})dz\right).\label{Atiyah-Manton-Skyrme-field}
\end{align}
Atiyah and Manton's proposal in \cite{AtiyahManton1989} was to consider (\ref{Atiyah-Manton-Skyrme-field}) as an ansatz for a Skyrme field in the case that the connection $D^A$ is a Yang-Mills instanton. An \textbf{instanton} is a connection $D^A$ whose curvature $F^A\in\Omega^2(\R^4,\su(2))$ is \textit{anti-self-dual}
\begin{align}\star_4F^A=-F^A,\label{ASD}\end{align}
and which leaves the Yang-Mills action
\begin{align}\label{YM-action}
    S_{YM}=-\int\tr(F^A\wedge\star_4 F^A)
\end{align}
finite (here we denote by $\star_4$ the Hodge-star on $\R^4$). In principal, any connection whose holonomy (\ref{Atiyah-Manton-Skyrme-field}) satisfies the correct boundary conditions ($U\to\mathbb{1}$ as $|\vec{x}|\to\infty$) could be considered, but instantons are preferable for many reasons. Firstly, instantons on $\R^4$ are geometrically equivalent to instantons (finite-action, anti-self-dual connections) on $S^4$. More precisely, every instanton on $\R^4$ is uniquely determined, via stereographic projection, by an instanton on $S^4$ \cite{Uhlenbeck1982}. In the limit $|\vec{x}|\to\infty$, the holonomy (\ref{Atiyah-Manton-Skyrme-field}) is taken around a constant loop in $S^4\cong\R^4\cup\{\infty\}$ (namely at the point $\infty$), and so $U\to\mathbb{1}$ in this limit, and thus the correct boundary conditions are satisfied. Moreover, any gauge transformation of $A$ acts on the holonomy $U$ via conjugation by a constant $SU(2)$ matrix. Such a transformation corresponds to an iso-rotation of the Skyrme field, which is a symmetry of the energy (\ref{Ordinary_Skyrme_energy}). Secondly, bundles over $S^4$ are characterised by their second Chern number $k=c_2(S^4)\in\Z$, referred to in this context as the \textbf{instanton number}. It can be shown that if $D^A$ has instanton number $k$, then the constructed Skyrme field (\ref{Atiyah-Manton-Skyrme-field}) has baryon number $\mathcal{B}=k$. Finally, another major benefit and motivation of using instantons in this ansatz is for studying low-energy interactions between nucleons. To do this within the Skyrme model, one needs to be able to choose a finite-dimensional manifold of Skyrme fields (with physically relevant coordinates). Instantons have a moduli space $\mathcal{I}_k$, which is known to be an $8k$-dimensional connected manifold \cite{atiyahhitchinsinger1978self}, where $k$ is the instanton number. Therefore, this schematic generates a connected ($8k-1$)-dimensional manifold of approximate charge $k$ Skyrme fields, given by $\bigslant{\mathcal{I}_k}{\R}$.

It turns out that Yang-Mills connections (and thus, instantons) are more than just convenient choices, but are in fact the natural connections to approximate Skyrme fields. This is explained by Sutcliffe's model \cite{sutcliffe2010skyrmions}, which we briefly review here. The key idea is to perform a `mode expansion' of the instanton gauge field in the $x^0$-direction, by expanding the fields in terms of a complete, orthonormal basis of $L^2(\R)$ -- the square integrable functions on $\R$. According to Sutcliffe, the correct choice of functions are the \textbf{Hermite functions} $\psi_n:\R\longrightarrow\R$, $n\in\N$, defined in \cite{AbramowitzStegun1964} by
\begin{align}\label{Hermite-func}
    \psi_n(x)=\dfrac{(-1)^n}{\sqrt{n!2^n\sqrt{\pi}}}e^{\frac{x^2}{2}}\dfrac{d^n}{dx^n}e^{-x^2}.
\end{align}
In a gauge where $A_\mu\to0$ as $|x^0|\to\infty$, we may perform a subsequent gauge transformation given by the inverse of
\begin{align}
    \Omega(x^0,\vec{x})=\mathcal{P}\exp\left(-\int_{-\infty}^{x^0}A_0(z,\vec{x})dz\right).
\end{align}
Under this, the instanton transforms such that $A_0=0$, and the remaining components satisfy the boundary condition $A_j\to U^{-1}\bdy_jU$ as $x^0\to+\infty$, where $U(\vec{x})=\Omega(\infty,\vec{x})$ is the holonomy. These components may be expanded in terms of the Hermite functions as
\begin{align}\label{mode-expansion-inst}
A_j=U^{-1}\bdy_jU\psi_+(x^0)+\sum_{n=0}^\infty V_j^n(\vec{x})\psi_n(x^0),
\end{align}
where $\psi_+:\R\longrightarrow\R$ is an additional basis function introduced in order to include the holonomy into the expansion. This is defined as
\begin{align}\label{psi_+}
\psi_+(x^0)=\dfrac{1}{\sqrt{2\sqrt{\pi}}}\int_{-\infty}^{x^0}\psi_0(z)dz=\dfrac{1}{2}+\dfrac{1}{\sqrt{\pi}}\int_{0}^{x^0/\sqrt{2}}e^{-w^2}dw,
\end{align}
and the normalisation given guarantees that $\psi_+(-\infty)=0$, and $\psi_+(\infty)=1$. The additional fields that appear in the expansion (\ref{mode-expansion-inst}) are the one-forms $V^n$, which physically are interpreted as (an infinite number of) \textit{vector mesons}.

The emergence of the Skyrme model from the Yang-Mills action is made apparent when the vector mesons in (\ref{mode-expansion-inst}) are artificially set to be $0$. After calculating the curvature $F^A$ of $D^A$ with respect to this truncation, one may perform the integration
$$-\frac{1}{2}\int\left(\int_{-\infty}^\infty\tr\left(\sum_{\mu,\nu}F^A_{\mu\nu}F^A_{\mu\nu}\right)dx^0\right)d^3x$$
corresponding to the contribution of the Yang-Mills lagrangian integrated along the $x^0$-axis. The resulting integral over $\R^3$ is precisely the energy for a static Skyrme model (\ref{Ordinary_Skyrme_energy}), with the coefficients $c_0$ and $c_1$ determined explicitly by integrating the Hermite function $\psi_+$ (\ref{psi_+}):
\begin{align}\label{c0-c1}
    c_0=\int_{-\infty}^\infty\left(\dfrac{d\psi_+}{dx^0}\right)^2dx^0=\dfrac{1}{2\sqrt{\pi}},\quad c_1=\int_{-\infty}^{\infty}\psi_+^2(1-\psi_+)^2dx^0\approx0.099.
\end{align}
\section{Skyrme models from periodic Yang-Mills}\label{sec.Sky-from-cal}
We will now proceed analogously as with Sutcliffe's construction to derive Skyrme models from Yang-Mills connections on $S^1\times\R^3$ (which later will be finite-action, and anti-self-dual, namely \textit{calorons}). Let $(t,\vec{x})$ be coordinates on $S^1\times\R^3$ with the flat product metric, identifying $t\sim t+\beta$, for some $\beta>0$. Let $A$ be an $SU(2)$ connection one form on $S^1\times\R^3$ with components $A_\mu$ satisfying the periodicity condition
\begin{align}
    A_\mu(t+\beta,\vec{x})=A_\mu(t,\vec{x}).
\end{align}
The parallel transport operator of $A$ along $S^1$ from $-\beta/2$ to $t$ is
\begin{align}\label{p-ordered-hol}
    \Omega(t,\vec{x})=\mathcal{P}\exp\left(-\int_{-\frac{\beta}{2}}^{t}{A_t(z,\vec{x})dz}\right),
\end{align}
where $\mathcal{P}$ denotes path-ordering. We may choose a gauge such that $\bdy_tA_t=0$, and in this gauge (\ref{p-ordered-hol}) becomes
\begin{align}
    \Omega(t,\vec{x})=\exp\left(-\left(t+\beta/2\right)A_t(\vec{x})\right).
\end{align}
The function $U:\R^3\longrightarrow SU(2)$ defined by $U(\vec{x})=\Omega(\beta/2,\vec{x})$, that is, the holonomy, is the candidate for a Skyrme field. We may further transform the connection $A$ with a non-periodic gauge transformation, given by the inverse of the parallel transport operator:
\begin{align}
A\mapsto\Omega^{-1} A \Omega+\Omega^{-1}d\Omega.
\end{align}
The transformed gauge potential satisfies
\begin{align}
    A_t&=0,\\
    A_j(t+\beta/2,\vec{x})&=U^{-1}A_j(t-\beta/2,\vec{x})U+U^{-1}\bdy_jU,\label{periodicity_gt}
\end{align}
that is, the $\R^3$ components are periodic up to a gauge transformation by $U^{-1}$. Now consider an $SU(2)$ connection one form $B$ on $\R^3$ defined via the transformed connection one-form $A$ as $B_j(\vec{x})=A_j(-\beta/2,\vec{x})$. Let $D^B=d+B$ be the connection covariant derivative defined by $B$. Then the boundary conditions above for the components $A_j$ imply
\begin{align}
\begin{array}{c}
    A_j(\beta/2,\vec{x})-A_j(-\beta/2,\vec{x})=U^{-1}D^B_jU,\\
    \bdy_t\left.\right|_{t=\beta/2}A_j=U^{-1}\bdy_t\left.\right|_{t=-\beta/2}A_jU.\end{array}\label{bdy_1}
\end{align}
We wish to perform a mode expansion of $A_j$ in an analogous way to (\ref{mode-expansion-inst}), in such a way that respects these boundary conditions. To do this, we shall consider a complete set of $L^2$-orthogonal functions $\{\varphi_+,\varphi_n\::\:n\in\N\}$, which span the space $L^2[-\beta/2,\beta/2]$, such that
$$\varphi_+(-\beta/2)=0,\quad\varphi_+(\beta/2)=1,\quad\text{and}\quad\varphi_n(\pm\beta/2)=0,\mbox{ }\forall\;n\in\N.$$
The function $\varphi_+$ is to be defined as
\begin{align}
    \varphi_+(t)=\dfrac{1}{\mathcal{N}}\int_{-\frac{\beta}{2}}^{t}\varphi_0(s)ds,
\end{align}
where $\mathcal{N}$ is chosen in such a way that $\varphi_+(\beta/2)=1$.

There are various choices that we can make here for the functions $\varphi$. One obvious choice would be for the mode expansion to be a Fourier expansion, that is, the basis is that of the trigonometric functions $(-1)^n\cos((n+1)\pi x/\beta)$. Another choice is to consider the \textit{ultra-spherical functions} $\phi_n^{(\alpha,\beta)}$ defined in \cite{AbramowitzStegun1964} by\footnote{For the purposes of later simplicity, we have changed the conventions in \cite{AbramowitzStegun1964} as $\alpha\mapsto2\alpha+\frac{1}{2}$.}
\begin{align}\label{ultraspherical}
    \phi_n^{(\alpha,\beta)}(x)=B_n^{(\alpha,\beta)}\left(1-\left(\frac{2x}{\beta}\right)^2\right)^{\alpha}C_n^\alpha\left(\frac{2x}{\beta}\right),
\end{align}
for all $\alpha>-\frac{1}{2}$, where
\begin{align*}
    B_n^{(\alpha,\beta)}&=\sqrt{\dfrac{2}{\beta}}\sqrt{\dfrac{n!(n+2\alpha+\frac{1}{2})\Gamma(2\alpha+\frac{1}{2})^2}{\pi 2^{-4\alpha}\Gamma(n+4\alpha+1)}},\quad\alpha\neq-\frac{1}{4},\quad B_n^{(-\frac{1}{4},\beta)}=\dfrac{n}{\sqrt{\pi\beta}},\\
    C_n^\alpha(x)&=\dfrac{(-1)^n}{2^nn!}\dfrac{\Gamma(2\alpha+1)\Gamma(n+4\alpha+1)}{\Gamma(4\alpha+1)\Gamma(n+2\alpha+1)}(1-x^2)^{-2\alpha}\dfrac{d^n}{dx^n}(1-x^2)^{n+2\alpha},
\end{align*}
with $\Gamma$ the usual gamma function
$$\Gamma(x)=\int_0^{\infty}{t^{x-1}e^{-t}dt}.$$
Ultimately we shall only be considering the ultra-spherical functions in our constructions, and not the Fourier expansion, for reasons outlined later in section \ref{section_inst-limit}.

Expanding $A_j$ in terms of any of these choices of functions, we obtain
\begin{align}
    A_j(t,\vec{x})=\varphi_+(t)L_j(\vec{x})+B_j(\vec{x})+\sum_{n=0}^\infty V_j^n(\vec{x})\varphi_n(t),\label{mode_expansion}
\end{align}
where the one form $L=U^{-1}D^BU$ is the left-invariant Maurer-Cartan current on $\R^3$, and the fields $V^n$ represent an infinite tower of vector mesons, in analogy to Sutcliffe's expansion of instantons (\ref{mode-expansion-inst}). In general, the one-forms $V^n$ may be subject to some additional constraints in order for (\ref{bdy_1}) to hold. If we set $V^n=0$ for all $n$, then the expanded gauge field (\ref{mode_expansion}) certainly satisfies the boundary conditions (\ref{bdy_1}) regardless of the choice of basis. The curvature $F^A$ of $A$ may be easily calculated in this basis in the case where $V^n=0$ as
\begin{align}
    F^A=-\dfrac{d}{dt}\varphi_+(t)L\wedge dt+\left(1-\varphi_+(t)\right)\left(F^B-\varphi_+(t)L\wedge L\right)+\varphi_+(t)U^{-1}F^BU,\label{curvature_expanded}
\end{align}
where $F^B=dB+B\wedge B$ is the curvature of $B$.

Recall that the Yang-Mills action for the connection $D^A$ is given by
$$S_{YM}=-\int{\tr(F^A\wedge\star_4 F^A)},$$
where here $\star_4$ denotes the Hodge-star on $S^1\times\R^3$. Using (\ref{curvature_expanded}), and integrating over the interval $[-\beta/2,\beta/2]$, we may write this as an energy functional over $\R^3$:
\begin{align}\label{gauged_Skyrme_energy}
    E=&\int\left(\lambda_0|L|^2+\lambda_1|L\wedge L|^2+\lambda_2|F^B|^2+\lambda_3\langle F^B ,U^{-1}F^BU\rangle\right.\\
    &\left.\qquad\qquad\qquad\quad-\lambda_4\left\langle F^B,L\wedge L\right\rangle-\lambda_5\left\langle U^{-1}F^BU,L\wedge L\right\rangle\right)d^3x.\notag
\end{align}
The coefficients $\lambda_p$ are determined by the $\varphi_+$ dependent integrals
\begin{align}\label{lambda_formulae}
\begin{array}{ll}
    \lambda_0=\int_{-\frac{\beta}{2}}^{\frac{\beta}{2}}\left(\dfrac{d\varphi_+}{dt}\right)^2dt,&\lambda_1=\int_{-\frac{\beta}{2}}^{\frac{\beta}{2}}\left(1-\varphi_+\right)^2\varphi_+^2dt,\\
    \lambda_2=\int_{-\frac{\beta}{2}}^{\frac{\beta}{2}}\left(1-2\varphi_++2\varphi_+^2\right)dt,&\lambda_3=2\int_{-\frac{\beta}{2}}^{\frac{\beta}{2}}\left(1-\varphi_+\right)\varphi_+dt,\\
    \lambda_4=2\int_{-\frac{\beta}{2}}^{\frac{\beta}{2}}\left(1-\varphi_+\right)^2\varphi_+dt,&\lambda_5=2\int_{-\frac{\beta}{2}}^{\frac{\beta}{2}}\left(1-\varphi_+\right)\varphi_+^2dt.
\end{array}
\end{align}
The energy (\ref{gauged_Skyrme_energy}) describes an $SU(2)$ Skyrme model on $\R^3$ coupled to a gauge field $B$, that is, a \textit{gauged Skyrme model}. Moreover, the holonomy transforms via gauge transformations of the caloron as
$$U(\vec{x})\mapsto g(\beta/2,\vec{x})U(\vec{x})g(-\beta/2,\vec{x})^{-1},$$
and when $g$ is $\beta$-periodic, this hence defines a gauge transformation $g(\vec{x}):=g(\beta/2,\vec{x})$ on $\R^3$. Coupled with the standard transformation
$$B\mapsto gBg^{-1}-dgg^{-1},$$
one can check that the energy (\ref{gauged_Skyrme_energy}) is invariant under all such gauge transformations induced by transforming the periodic connection $D^A$.

There have been various previous considerations of static gauged Skyrme models \cite{ArthurTchrakian1996GaugedSky,BrihayeHartmannTchhrakian2001MonopolesGaugeSky,BrihayeHillZachos2004BoundingGaugeSkyMass,PietteTchrakian2000static}. In the cases where the gauge field takes values in $\su(2)$ \cite{ArthurTchrakian1996GaugedSky,BrihayeHartmannTchhrakian2001MonopolesGaugeSky}, the terms in the energies considered are $|L|^2$, $|L\wedge L|^2$, and $|F|^2$. The model (\ref{gauged_Skyrme_energy}) that we have derived contains additional `cross terms' which have not been considered in the context of any gauged Skyrme model before, but nevertheless are gauge-invariant, and seem to be natural terms to include due to their appearance from this construction.

\subsection{A family of gauged Skyrme energies}\label{section_family-G-Sky-en}
Recall the ultraspherical functions $\phi_n^{(\alpha,\beta)}$ defined for $\alpha>-1/2$ and $\beta>0$ by (\ref{ultraspherical}). We shall now consider the expansion (\ref{mode_expansion}) in terms of these functions. Since we are only interested at this time in the case where $V^n=0$, the only function that contributes to the energy (\ref{gauged_Skyrme_energy}) is the additional function $\phi_+^{(\alpha,\beta)}$, given by
\begin{align}
    \phi_+^{(\alpha,\beta)}(t)&=\dfrac{2^{\frac{1}{2}-2\alpha}\Gamma\left(\alpha+\frac{3}{2}\right)}{\Gamma\left(\alpha+1\right)\sqrt{(2\alpha+\frac{1}{2})\beta}}\sqrt{\dfrac{\Gamma(4\alpha+1)}{\Gamma(2\alpha+\frac{1}{2})^2}}\int_{-\frac{\beta}{2}}^{t}{\phi_0^{(\alpha,\beta)}(z)dz}\notag\\
    &=\dfrac{1}{2}+\dfrac{2t}{\beta}\dfrac{\Gamma(\alpha+\frac{3}{2})}{\sqrt{\pi}\Gamma(\alpha+1)}\;{}_2F_1\left(\frac{1}{2},-\alpha;\frac{3}{2};\frac{4t^2}{\beta^2}\right).\label{phi_+}
\end{align}
Here ${}_2F_1(a,b;c;z)$ denotes the hypergeometric function  \cite{AbramowitzStegun1964}, which belongs to the family of \textit{generalised hypergeometric functions}, defined for $p,q\in\N$, $p\leq q+1$, by the series\footnote{In general, if $p=q+1$, the numbers $a_r,b_r\in\R$ are subject to strict constraints in order for the series to converge. In our case these are satisfied, but in general see \cite{AbramowitzStegun1964}.}
\begin{align}\label{hyper-geometric-function}
_pF_q(a_1,\dots,a_p;b_1,\dots,b_q;x)=\dfrac{\Gamma(b_1)\cdots\Gamma(b_q)}{\Gamma(a_1)\cdots\Gamma(a_p)}\sum_{n=0}^\infty{\dfrac{\Gamma(a_1+n)\cdots\Gamma(a_p+n)}{\Gamma(b_1+n)\cdots\Gamma(b_q+n)}}\dfrac{x^n}{n!}.   
\end{align}
The normalisation in (\ref{phi_+}) has been chosen so that $\phi_+^{(\alpha,\beta)}(-\beta/2)=0$, and $\phi_+^{(\alpha,\beta)}(\beta/2)=1$, and these identities may be straightforwardly checked by utilising the integral formula for the `beta function'
\begin{align}\label{beta_int}
    \dfrac{\Gamma(z)\Gamma(w)}{\Gamma(z+w)}=2\int_{0}^{\frac{\pi}{2}}\sin^{2z-1}(x)\cos^{2w-1}(x)dx,\quad\text{for }\Re(z),\Re(w)>0.
\end{align}
For $\beta=1$, the function (\ref{phi_+}) has a power series expansion given by
\begin{align}\label{phi_+-series}
\phi_+^{(\alpha,1)}(t)=\dfrac{1}{2}+\dfrac{\Gamma(\alpha+\frac{3}{2})}{\sqrt{\pi}\Gamma(\alpha+1)}\sum_{k=0}^{\infty}\dfrac{1}{2k+1}\dfrac{(-1)^k}{k!}\left(\prod_{r=0}^{k-1}(\alpha-r)\right)(2t)^{2k+1}.
\end{align}
It is easy to calculate that this series converges for all $|t|<\frac{1}{2}$, and since $\alpha>-\frac{1}{2}$, its value at $t=\pm\frac{1}{2}$ is well-defined. So since $\phi_+^{(\alpha,\beta)}(t)=\phi_+^{(\alpha,1)}(t/\beta)$, this means the function $\phi_+^{(\alpha,\beta)}$ is well-defined, and clearly continuous on $[-\frac{\beta}{2},\frac{\beta}{2}]$, for all $\alpha>-\frac{1}{2}$, and $\beta>0$.

The next step is to compute the coefficients $\lambda_p$ appearing in the energy (\ref{gauged_Skyrme_energy}). These are determined by the formulae (\ref{lambda_formulae}) and require the evaluation of the integrals
\begin{align}\label{I_r}
    I_r(\alpha)=\int_{-\frac{1}{2}}^{\frac{1}{2}}\left(\phi_+^{(\alpha,1)}(t)\right)^rdt,
\end{align}
for all $r=0,\dots,4$. For $r=0,1$, the integrals $I_r$ are easy to calculate, namely
\begin{align}\label{I_0,1}
    I_0(\alpha)=1,\quad
    I_1(\alpha)=\frac{1}{2}.
\end{align}
After a significant amount of calculation, which we reserve to appendix \ref{appendix-integrals}, we have
\begin{align}
    I_2(\alpha)=\dfrac{1}{2}\left(1-\dfrac{\Gamma(2\alpha+2)\Gamma(\alpha+\frac{3}{2})^2}{\Gamma(2\alpha+\frac{5}{2})\Gamma(\alpha+1)\Gamma(\alpha+2)\sqrt{\pi}}\right).\label{I_2}
\end{align}
The integral $I_3$ is straightforwardly seen to be determined by $I_2$, namely via the formula
\begin{align}
    I_3(\alpha)=\dfrac{3}{2}I_2(\alpha)-\dfrac{1}{4}.\label{I_3}
    \end{align}
Finally, evaluating $I_4$ in general is a much harder problem, and we do not have an explicit formula in terms of elementary functions of $\alpha$, however a partial formula is presented in appendix \ref{appendix-integrals}. Nevertheless, we may use (\ref{gauged_Skyrme_energy}) and (\ref{lambda_formulae}), along with (\ref{I_0,1}) and (\ref{I_3}), to formulate a family of gauged Skyrme energies
\begin{align}\label{family-gauged-energies-beta}
    E_{\alpha,\beta}&=\int\left(\dfrac{\kappa_0}{\beta}|L|^2+\dfrac{\beta}{2}\kappa_1|L\wedge L|^2+\beta|F^B|^2\right.\\
    &\left.+\beta\kappa_2\left(\left\langle U^{-1}F^BU,F^B\right\rangle-\dfrac{1}{2}\left\langle F^B+U^{-1}F^BU,L\wedge L\right\rangle-|F^B|^2\right)\right)d^3x,\notag
\end{align}
where we have introduced the notation
\begin{align}
    \kappa_0(\alpha)&=\dfrac{2}{\sqrt{\pi}}\dfrac{\Gamma(2\alpha+1)\Gamma(\alpha+\frac{3}{2})^2}{\Gamma(2\alpha+\frac{3}{2})\Gamma(\alpha+1)^2},\label{kappa_0}\\
    \kappa_1(\alpha)&=1+2I_4-4I_2,\label{kappa_1}\\
    \kappa_2(\alpha)&=1-2I_2.\label{kappa_2}
\end{align}
\subsection{The instanton/weak coupling limit}\label{section_inst-limit}
Expanding the caloron gauge field in terms of the complete, orthonormal basis of $L^2([-\beta/2,\beta/2])$, given by the ultra-spherical functions, has revealed a family of gauged Skyrme energies parameterised by the period $\beta>0$ of the caloron, and the ultraspherical parameter $\alpha>-1/2$. Other more complicated models may also be obtained by including some or all of the vector mesons $V^n$ in (\ref{mode_expansion}), resulting in extensions to the family of energies we already have. This choice of functions already has an advantage over considering a simpler basis of $L^2([-\beta/2,\beta/2])$, for example the trigonometric functions, since we may vary the parameter $\alpha$ to explore different properties of the energy (\ref{family-gauged-energies-beta}). Another advantage of this choice is the relationship between the ultraspherical functions and the Hermite functions (\ref{Hermite-func}) and (\ref{psi_+}). Indeed, consider a limit where $\alpha,\beta\to\infty$, such that $\alpha/\beta^2\to1/8$. In this limit, the weight function for the ultraspherical functions satisfies
\begin{align}\label{weight-limit}
    \left(1-\left(\dfrac{2x}{\beta}\right)^2\right)^\alpha\longrightarrow e^{-\frac{x^2}{2}}.
\end{align}
Also, for all $z>0$ sufficiently large, we have \cite{AbramowitzStegun1964} that for all $a>0$, $b\in\R$
\begin{align}\label{Gamma-function-asymptotics}
    \Gamma(az+b)\sim\sqrt{2\pi}e^{-az}(az)^{az+b-\frac{1}{2}}.
\end{align}
From the formula (\ref{ultraspherical}), and using both (\ref{weight-limit}) and (\ref{Gamma-function-asymptotics}), we therefore have that for $\alpha,\beta>0$ large:
\begin{align*}
    \phi_n^{(\alpha,\beta)}(x)&\sim(-1)^n\sqrt{\dfrac{2^{\frac{3}{2}-2n}}{\beta n!\sqrt{\pi}\alpha^{n-\frac{1}{2}}}}\left(1-\left(\frac{2x}{\beta}\right)^2\right)^{-\alpha}\left(\dfrac{\beta}{2}\right)^n\dfrac{d^n}{dx^n}\left(1-\left(\frac{2x}{\beta}\right)^2\right)^{2\alpha}\\
    &=(-1)^n\sqrt{\dfrac{(\beta^2/8\alpha)^{n-\frac{1}{2}}}{n!2^n\sqrt{\pi}}}\left(1-\left(\frac{2x}{\beta}\right)^2\right)^{-\alpha}\dfrac{d^n}{dx^n}\left(1-\left(\frac{2x}{\beta}\right)^2\right)^{2\alpha}\\
    &\xrightarrow{\alpha,\beta\to\infty}\dfrac{(-1)^n}{\sqrt{n!2^n\sqrt{\pi}}}e^{\frac{x^2}{2}}\dfrac{d^n}{dx^n}e^{-x^2},
\end{align*}
which is the formula (\ref{Hermite-func}) for the Hermite functions $\psi_n$. As $\psi_+$ and $\phi_+^{(\alpha,\beta)}$ are defined as normalised integrals (over $\R$ and $[-\beta/2,\beta/2]$) of $\psi_0$ and $\phi_0^{(\alpha,\beta)}$ respectively, it hence follows that the limit $\phi_+^{(\alpha,\beta)}\to\psi_+$ also holds.

The main consequence of this limiting behaviour is that any gauged Skyrme model derived from the ultraspherical functions in the mode expansion (\ref{mode_expansion}) of a caloron (with any number of vector mesons included) reduces, in a particular limit as $\alpha,\beta\to\infty$, to the Sutcliffe model derived from an instanton mode expansion (\ref{mode-expansion-inst}), with the same number of vector mesons included. In particular, in the case that $V^n=0$, we have that the energy for an ordinary Skyrme model is made manifest in this limit as a part of the gauged Skyrme energy (\ref{family-gauged-energies-beta}). Since the limit $\beta\to\infty$ corresponds to an infinitely periodic caloron, which may in many cases be recognised as an instanton on $\R^4$, we shall hence call the limit $\alpha,\beta\to\infty$ with $\alpha/\beta^2\to1/8$ the \textit{instanton} or \textit{weak coupling limit} of the gauged Skyrme energy (\ref{family-gauged-energies-beta}).
\subsection{Scaling and parameter fixing}
It is straightforward to show that under a re-scaling of the spatial coordinates via $\vec{x}\mapsto\frac{1}{\beta}\vec{x}$, the energy (\ref{family-gauged-energies-beta}) transforms as
$$E_{\alpha,\beta}\mapsto\dfrac{1}{\beta^3}E_{\alpha,1}.$$
What this means is that as far as the functional $E_{\alpha,\beta}$ is concerned, the parameter $\beta$ only affects it up to a re-scaling of the energy and length units. Therefore, in order to make things simpler, we may without loss of generality choose to set $\beta=1$, which we shall do from now on. For notational brevity, we shall also introduce the notation $E_{\alpha}=E_{\alpha,1}$ and henceforth consider the energies
\begin{align}\label{family-gauged-energies}
    E_{\alpha}&=\int\left(\kappa_0|L|^2+\dfrac{\kappa_1}{2}|L\wedge L|^2+|F^B|^2\right.\\
    &\left.+\kappa_2\left(\left\langle U^{-1}F^BU,F^B\right\rangle-\dfrac{1}{2}\left\langle F^B+U^{-1}F^BU,L\wedge L\right\rangle-|F^B|^2\right)\right)d^3x.\notag
\end{align}
\subsection{Topological charge and energy bounds}\label{section_top-en-b}
The topological charge for a Yang-Mills connection is given by the formula
\begin{align}\label{YM-top-charge}
    Q_{YM}=\dfrac{1}{8\pi^2}\int\tr\left(F^A\wedge F^A\right).
\end{align}
Similarly to the energy, we may calculate the topological charge for our gauged Skyrme model (\ref{family-gauged-energies-beta}) by inserting the expansion (\ref{mode_expansion}) with $V^n=0$ into (\ref{YM-top-charge}) and integrating over $[-\beta/2,\beta/2]$. We hence obtain the formula
\begin{align}\label{top-charge_gauged-Skyrme}
    \mathcal{Q}&=\dfrac{1}{8\pi^2}\int{\tr\left(\dfrac{1}{3}L\wedge L\wedge L-L\wedge\left(F^B+U^{-1}F^BU\right)\right)},
\end{align}
which is precisely the usual topological charge for a gauged Skyrme model \cite{ArthurTchrakian1996GaugedSky,BrihayeHartmannTchhrakian2001MonopolesGaugeSky,PietteTchrakian2000static}, and reduces to the topological charge (\ref{baryon-number_integral}) for the ordinary Skyrme model when $B=0$. A straightforward argument shows that the Yang-Mills action (\ref{YM-action}) has the topological bound
$$S_{YM}\geq 8\pi^2|Q_{YM}|,$$
which is saturated by anti-self-dual connections. This may be applied in the context of our gauged Skyrme models, and we immediately have the energy bound
\begin{align}\label{YM_Skyrme-bound}
    E_\alpha\geq8\pi^2|\mathcal{Q}|
\end{align}
for all $\alpha>-\frac{1}{2}$. For analogous reasons to why ordinary skyrmions cannot attain the energy bound (\ref{Fadeev-bound}), any minimisers of (\ref{family-gauged-energies-beta}) will not attain the energy bound (\ref{YM_Skyrme-bound}) either. However, the model which includes all of the vector mesons will be BPS, in as much as minimisers can obtain the topological bound (\ref{YM_Skyrme-bound}), simply because they are a completion of a caloron mode expansion, and calorons \textit{are} BPS. In particular, calorons generate exact BPS solutions to this fully extended model.

For the model which we are concerned with, that is, the one with no vector mesons, described by the family of energies (\ref{family-gauged-energies-beta}), there is no reason why the bound (\ref{YM_Skyrme-bound}) is the best bound that can be found for all $\alpha>-\frac{1}{2}$. In order to make an attempt at a stronger bound, we consider the following quadratic forms on $\R^4$:
\begin{align}
    \omega_E&=\kappa_0x^2+\frac{\kappa_1}{2}y^2+I_2z^2+I_2w^2+\kappa_2zw-\frac{\kappa_2}{2}y(w+z),\label{omega_E}\\
    \omega_{\mathcal{Q}}&=\dfrac{1}{3}xy-x(w+z).\label{omega_Q}
\end{align}
\begin{lemma}\label{lem-pos-def}
The quadratic forms $\omega_E$ and $\omega_{\mathcal{Q}}$, given by (\ref{omega_E}) and (\ref{omega_Q}) respectively, have signatures $\{+,+,+,+\}$ and $\{+,-,0,0\}$ respectively, for all $\alpha>-\frac{1}{2}$.
\end{lemma}
\textit{Proof}. The signature of $\omega_{\mathcal{Q}}$ is easily determined by calculating the eigenvalues of the associated symmetric matrix, which are $\wt{\lambda}_{\pm}=\pm\sqrt{19}/6$ and $\wt{\lambda}_0=0$, with multiplicities $1$, $1$, and $2$ respectively. In contrast, the signature of $\omega_E$ depends on the coefficients $\kappa_j$. Indeed, it is straightforward to show that the associated symmetric matrix to $\omega_E$ has eigenvalues
\begin{align}
    \lambda_0&=\kappa_0,\quad\lambda_1=\frac{1}{2}-\kappa_2,\\
    \lambda_{\pm}&=\frac{1}{4}\left(\kappa_1+1\pm\sqrt{(\kappa_1-1)^2+2\kappa_2^2}\right).
\end{align}
Since $\alpha>-\frac{1}{2}$, we have by the formula (\ref{kappa_0}) that $\lambda_0>0$. It is straightforward to see from (\ref{phi_+}) that
\begin{align}\label{eta-bound}|\eta_\alpha(t)|\leq\frac{1}{2}\quad\forall\:t\in\left[-\frac{1}{2},\frac{1}{2}\right],\:\forall\:\alpha>-\frac{1}{2},
\end{align}
where have defined $\eta_\alpha(t)=\phi_+^{(\alpha,1)}(t)-\frac{1}{2}$, which in particular is an odd function. We also have that
\begin{align}
I_2&=\int_{-\frac{1}{2}}^{\frac{1}{2}}\left(\frac{1}{2}+\eta_\alpha(t)\right)^2dt=\frac{1}{4}+\int_{-\frac{1}{2}}^{\frac{1}{2}}\eta_\alpha(t)^2dt,\label{I2}\\
I_4&=\int_{-\frac{1}{2}}^{\frac{1}{2}}\left(\frac{1}{2}+\eta_\alpha(t)\right)^4dt=\frac{1}{16}+\int_{-\frac{1}{2}}^{\frac{1}{2}}\left(\frac{3}{2}\eta_\alpha(t)^2+\eta_\alpha(t)^4\right)dt.\label{I4}
\end{align}
It hence follows from (\ref{kappa_1})-(\ref{kappa_2}) and (\ref{I2})-(\ref{I4}) that
\begin{align}
    \kappa_1&=\dfrac{1}{8}+\int_{-\frac{1}{2}}^{\frac{1}{2}}\left(2\eta_\alpha(t)^4-\eta_\alpha(t)^2\right)dt,\label{kappa_1-int}\\
    \kappa_2&=\dfrac{1}{2}-2\int_{-\frac{1}{2}}^{\frac{1}{2}}\eta_\alpha(t)^2dt.\label{kappa_2-int}
\end{align}
Combining (\ref{eta-bound}) with (\ref{kappa_1-int})-(\ref{kappa_2-int}), we may determine
\begin{align}
    -\frac{1}{8}<\kappa_1<\dfrac{1}{4},\quad0\leq\kappa_2<\frac{1}{2}.
\end{align}
From these inequalities we may conclude that $\lambda_1,\lambda_+>0$ for all $\alpha>-\frac{1}{2}$. Finally, the condition $\lambda_{-}>0$ is equivalent to the inequality
\begin{align}\label{kappas_cond}
    \kappa_2^2<2\kappa_1.
\end{align}
To prove (\ref{kappas_cond}), we note by (\ref{kappa_1-int})-(\ref{kappa_2-int}), we have
\begin{align*}
    2\kappa_1&=\dfrac{1}{4}+2\int_{-\frac{1}{2}}^{\frac{1}{2}}\left(2\eta_\alpha(t)^4-\eta_\alpha(t)^2\right)dt,\\
    \kappa_2^2&=\dfrac{1}{4}+4\left(\int_{-\frac{1}{2}}^{\frac{1}{2}}\eta_\alpha(t)^2dt\right)^2-2\int_{-\frac{1}{2}}^{\frac{1}{2}}\eta_\alpha(t)^2dt.
\end{align*}
Thus, (\ref{kappas_cond}) is equivalent to the inequality
\begin{align*}
    \left(\int_{-\frac{1}{2}}^{\frac{1}{2}}\eta_\alpha(t)^2dt\right)^2<\int_{-\frac{1}{2}}^{\frac{1}{2}}\eta_\alpha(t)^4dt,
\end{align*}
which is simply the Cauchy-Schwarz inequality. Thus, all eigenvalues are strictly positive.
\hfill$\square$\\

We may now prove the existence of the following energy bound.
\begin{theorem}
The gauged Skyrme energy $E_\alpha$ satisfies the topological bound
\begin{align}
    E_\alpha\geq 8\pi^2C(\alpha)|\mathcal{Q}|,\label{optimal-bound}
\end{align}
where $C(\alpha)$ is given by
\begin{align}\label{C-alpha}
    C(\alpha)=\sqrt{\dfrac{9\kappa_0\left(2\kappa_1-\kappa_2^2\right)}{1+ 18\kappa_1-6\kappa_2}},
\end{align}
and this bound is the best bound that may be obtained by completing the square.
\end{theorem}
\textit{Proof}. Before we prove the existence of the bound (\ref{optimal-bound}), we shall first confirm that $C(\alpha)$ given by (\ref{C-alpha}) is real. Indeed, we clearly have $\kappa_0>0$ by (\ref{kappa_0}), and (\ref{kappas_cond}) shows that the numerator of $C(\alpha)^2$ is positive. For the denominator, we have by (\ref{kappas_cond}) that
$$1+18\kappa_1-6\kappa_2>(3\kappa_2-1)^2\geq0.$$
Thus comparing this with (\ref{C-alpha}), we see that $C(\alpha)^2>0$, so $C(\alpha)\in\R$. Now consider the quadratic form
$$\omega_t=(1-t)\omega_E-t\omega_\mathcal{Q},$$
for $t\in[0,1]$. The eigenvalues of the associated symmetric matrix to $\omega_t$ are real and depend continuously on $t\in[0,1]$. Since $\omega_0=\omega_{E}$ and $\omega_1=-\omega_{\mathcal{Q}}$, we may apply lemma \ref{lem-pos-def} and the intermediate value theorem to deduce the existence of $s\in(0,1)$ such that the matrix associated to $\omega_s$ has a zero eigenvalue. Now, it is straightforward to see from (\ref{family-gauged-energies}) and (\ref{top-charge_gauged-Skyrme}) that determining an optimal bound of the form (\ref{optimal-bound}) is equivalent to finding the maximal value of $C$ such that the quadratic form
\begin{align}\label{omega_c}
    \Omega_C=\omega_E-C\omega_\mathcal{Q}
\end{align}
is non-negative. In turn, this is equivalent to the quadratic form $\omega_t$ being non-negative for some $t\in[0,1)$. The associated symmetric matrix to the quadratic form (\ref{omega_c}) has characteristic polynomial
\begin{align*}
   \chi(\Omega_C)(\lambda)=\dfrac{1}{144}\left(2 \kappa_2-1 + 2 \lambda\right)&\left(C^2\left(1+ 18\kappa_1-6\kappa_2-38\lambda\right)\right.\\
   &\left.+ 
   9 (\kappa_0 - \lambda)(\kappa_2^2-2\kappa_1+4(1+\kappa_1)\lambda-8\lambda^2))\right).
\end{align*}
The signature of $\Omega_C$ is given by the signs of the roots of $\chi(\Omega_C)$, which are $\lambda_1=\frac{1}{2}(1-2\kappa_2)$, and $\lambda_{2,3,4}$ determined by the roots of the polynomial $P_C$:
\begin{align*}
    P_C(\lambda)&=C^2\left(1+ 18\kappa_1-6\kappa_2-38\lambda\right)+9(\kappa_0-\lambda)(\kappa_2^2-2\kappa_1+4(1+\kappa_1)\lambda-8\lambda^2).
\end{align*}
In particular, $P_C$ has the root $\lambda=0$ if and only if $C=C_{\mathrm{max}}$ where
\begin{align}\label{C_max}
C_{\mathrm{max}}^2=\dfrac{9\kappa_0\left(2\kappa_1-\kappa_2^2\right)}{1+ 18\kappa_1-6\kappa_2}.
\end{align}
The value $s\in(0,1)$ such that $\omega_s$ has a zero eigenvalue is determined by asking that the polynomial $P_{\frac{s}{1-s}}$ has zero as a root, and it hence follows from the above argument that this is uniquely given by
$$s=\dfrac{C_{\mathrm{max}}}{1+C_{\mathrm{max}}},$$
where $C_{\mathrm{max}}$ here is chosen as the positive root of (\ref{C_max}), namely $C(\alpha)$ given in (\ref{C-alpha}). Since this value $s\in(0,1)$ is unique, we must also have that $\omega_t$ is positive definite for all $0\leq t<s$, and of mixed signature for all $s<t\leq1$, meaning that (\ref{optimal-bound}) is optimal.\hfill$\square$\\

\begin{figure}[ht]
\centering
\includegraphics[width=0.8\textwidth]{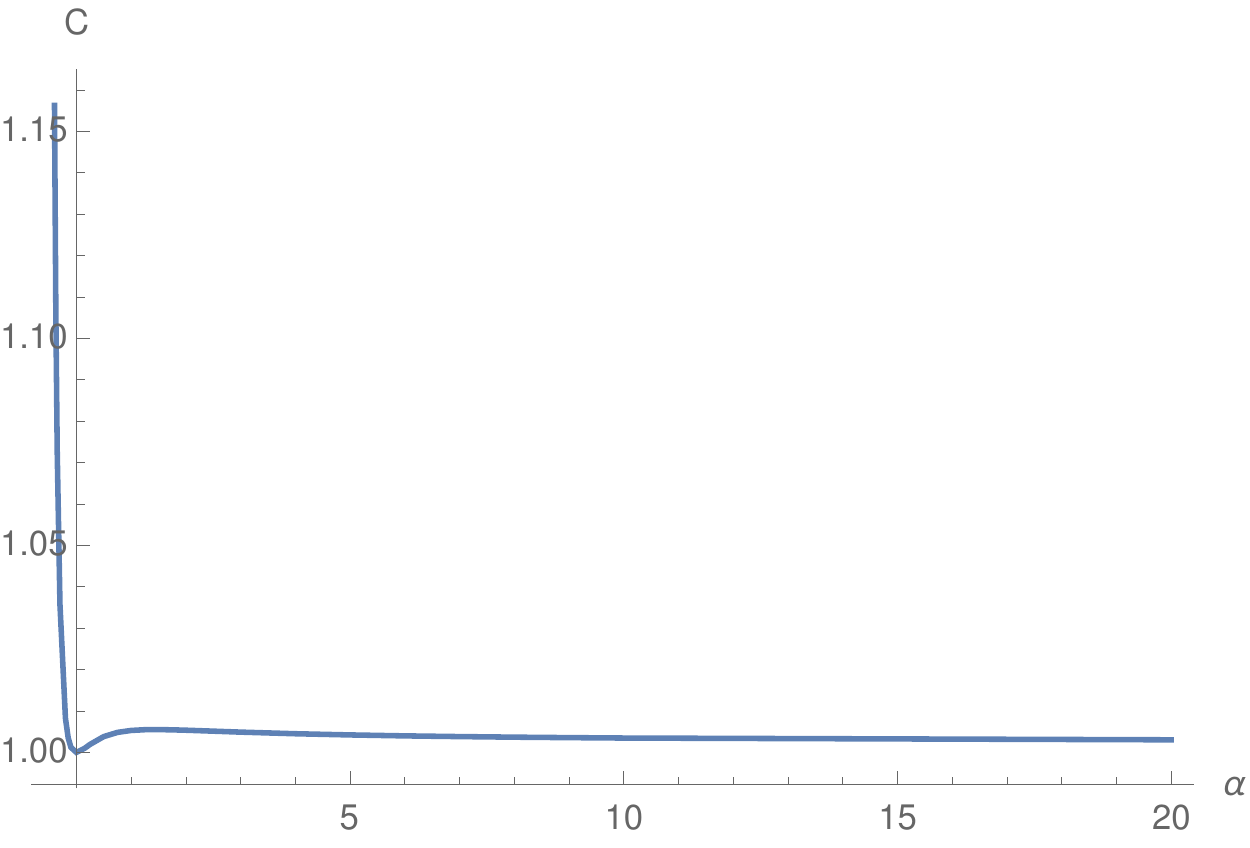}
\caption{The function $C(\alpha)$ appearing in the topological bound $E_\alpha\geq 8\pi^2C(\alpha)|\mathcal{Q}|$.}
\label{fig.top_bound_plot}
\end{figure}

In figure \ref{fig.top_bound_plot}, we plot numerically the function $C(\alpha)$ given in (\ref{C-alpha}).
For $\alpha\geq0$, the bound is observed to be relatively stable, with $C\sim1$ for all $\alpha$, whereas, for $\alpha<0$, the bound becomes extremely large -- whilst it cannot be seen in the plot, we found that $C(-0.499)\approx8.29811$. This suggests that the model is far more well-behaved for $\alpha\geq0$. Due to the analysis in section \ref{section_inst-limit}, from the formulae (\ref{Fadeev-bound}) and (\ref{c0-c1}), we expect the limit
\begin{align}
    \lim_{\alpha\to\infty}C(\alpha)=6\sqrt{c_0c_1}\approx1.0027
\end{align}
to hold. Interestingly, the minimum value of $C$ that we have found is given uniquely by $\alpha=0$, where $C(0)=1$, i.e. the topological bound for $\alpha=0$ agrees with the Yang-Mills bound, whereas in all other cases, we have a stronger bound.
\section{Gauged skyrmions and their caloron approximations}\label{sec.numerics}
Calorons are finite-action, anti-self-dual connections on $S^1\times\R^3$, and are therefore natural candidates for approximating gauged skyrmions via their holonomies in the $S^1$ direction. The boundary conditions for calorons in general are extremely subtle, and we shall not attempt to summarise them all here. For this we direct the reader to the PhD thesis of Nye \cite{Nyethesis}. A key point (which we introduce here mostly for the purpose of notation) is that $SU(2)$ calorons exhibit two topological charges $k,m\in\Z$ called the \textit{instanton number} and \textit{magnetic charge} in direct analogy with instantons on $\R^4$ and monopoles on $\R^3$. Calorons are often referred to by their charges as `$(k+m,k)$-calorons'.

There are two types of calorons which we shall concentrate on. Firstly are the $(m,0)$-calorons. These are simply given by monopoles on $\R^3$. A \textbf{monopole} is a pair $(\wt{A},\Phi)$ comprising of an $SU(2)$ connection $D^{\wt{A}}$, and Higgs field $\Phi:\R^3\longrightarrow\su(2)$, satisfying the Bogomolny equation $\star_3D^{\wt{A}}\Phi=-F^{\wt{A}}$, and the boundary condition
\begin{align}\label{bdy-monopoles}
    -\dfrac{1}{2}\tr(\Phi^2)\longrightarrow\nu^2,\quad\text{as }|\vec{x}|\to\infty,
\end{align}
where $\nu\in(0,\pi/\beta]$. Here $\beta$ is the period of the caloron (which we shall, as explained before, also take to be $\beta=1$). The magnetic charge $m\in\Z$ emerges as a consequence of the boundary condition (\ref{bdy-monopoles}) which says that the Higgs field at infinity takes values in a $2$-sphere of radius $\nu$ in $\su(2)$, that is, it is a map $\Phi_\infty:S^2_\infty\longrightarrow S^2_{\nu}\subset\su(2)$. The magnetic charge $m\in\Z$ is the degree of this map. An $(m,0)$-caloron is obtained from a monopole by setting $A_t=\Phi$, and $A_j=\wt{A}_j$ for $j=1,2,3$. The associated Skyrme field $U$ generated by the holonomy of a monopole satisfies the boundary condition
\begin{align}\label{Skyrme-monopole-bdy-cond}
    \dfrac{1}{2}\tr U\longrightarrow\cos\nu.
\end{align}
One can show that when $\beta=1$, the topological charge for a monopole, and hence the associated gauged Skyrme field, is $\mathcal{Q}=\frac{m\nu}{\pi}$. This example highlights a key difference between gauged skyrmions and ordinary skyrmions: the Skyrme field for a gauged Skyrme model does not have to satisfy the boundary condition $U\to\mathrm{constant}$ as $r\to\infty$ in order for it to have a well-defined topological charge, and moreover, the whilst the baryon number of a skyrmion is always an integer, the gauged skyrmion generated by a monopole can have a charge given by any real number.

The second type of caloron we shall be concerned with are those with $0$ magnetic charge, that is the $(k,k)$-calorons. These satisfy (among others) the boundary condition in a local gauge near $|\vec{x}|=\infty$
\begin{align}
    A_t\sim\begin{pmatrix}
    \imath\mu&0\\
    0&-\imath\mu\end{pmatrix}+O(|\vec{x}|^{-2}),
\end{align}
where $\mu\in[0,\pi/\beta]$. This boundary condition implies that the holonomy
$$U(\vec{x})=\mathcal{P}\exp\left(-\int_{S^1}A_tdt\right),$$
and hence the associated Skyrme field, satisfies the boundary condition
\begin{align}\label{Skyrme-inst-bdy}
    U\longrightarrow\begin{pmatrix}
    e^{-\imath\mu}&0\\
    0&e^{\imath\mu}\end{pmatrix},\quad\text{as }|\vec{x}|\to\infty,
\end{align}
and so $U$ has a well-defined degree, and that is precisely the instanton number $k\in\Z$. The topological charge of a $(k,k)$-caloron and its associated gauged skyrmion is $\mathcal{Q}=k$. According to the monad descriptions of calorons and instantons \cite{CharbonneauHurtubise2008Rat.map,donaldson1984instantons}, the moduli space of $(k,k)$-calorons is embedded in the moduli space of $k$-instantons, furthermore, various examples of instantons are exhibited in the limit $\beta\to\infty$ of $(k,k)$-calorons \cite{Harland2007,HarringtonShepard1978periodic,Ward2004}. For these reasons, the $(k,k)$-calorons are the most `instanton-like' calorons.
\subsection{Hedgehogs}
The group $O(3)$ of spherical rotations and reflections has a natural action on the space of gauged Skyrme configurations given by
\begin{align}
    \mathfrak{R}\cdot(U,B)&=\left(U\circ \mathfrak{R},\mathfrak{R}^\ast B\right),\label{SO(3)-action-gauge-Skyrme}\\
    \sigma\cdot(U,B)&=\left((U\circ\sigma)^{-1},\sigma^\ast B\right),\label{chiral-reflection}
\end{align}
where $\mathfrak{R}:\R^3\longrightarrow\R^3$ represents an element of $SO(3)$, and $\sigma:\R^3\to\R^3$ is the parity transformation $\vec{x}\mapsto-\vec{x}$. A gauged skyrmion $(U,B)$ is called $H$\textbf{-symmetric} if for all $h\in H\subset O(3)$, it is invariant under the relevant actions (\ref{SO(3)-action-gauge-Skyrme})-(\ref{chiral-reflection}), up to gauge transformations. These actions combined leave the energy (\ref{family-gauged-energies}) invariant, so are symmetries of the field theory.

We shall concentrate on the most symmetric case, namely full spherical symmetry. It is well-known that the most general representative $(U,B)$ of a gauged Skyrme configuration which is $O(3)$-symmetric is given by the fields in the \textit{hedgehog ansatz}
\begin{align}\label{hedgehog_ansatz}
    U=\exp\left(\imath f(r)\dfrac{\vec{x}\cdot\vec{\sigma}}{r}\right),\quad B=\dfrac{\imath}{2}\left(g(r)-1\right)\dfrac{\vec{x}\times\vec{\sigma}}{r^2}\cdot d\vec{x},
\end{align}
where $f,g:(0,\infty)\longrightarrow\R$ are functions in the radial direction $r=|\vec{x}|$. Imposing this spherically symmetric form, the family of energies (\ref{family-gauged-energies}) reduce to the one-dimensional integral
\begin{align}\label{Energy_seq_hedgehog}
    E_\alpha^H=&8\pi\int_{0}^\infty\left(\kappa_0\left(r^2{f'}^2+2g^2\sin^2f\right)+\dfrac{1}{4r^2}\left(1-g^2\right)^2+\dfrac{{g'}^2}{2}\right.\notag\\
    &\left.\qquad\qquad+\kappa_2\left(\sin^2f\left(\dfrac{g^2}{r^2}\left(1-g^2\right)-{g'}^2\right)-f'gg'\sin2f\right)\right.\\
    &\left.\qquad\qquad\qquad\qquad+2\kappa_1\left(\dfrac{g^4}{r^2}\sin^4f+2{f'}^2g^2\sin^2f\right)\right)dr.\notag
\end{align}
Similarly, we may calculate the topological charge for the fields in the hedgehog ansatz as
\begin{align}
    \mathcal{Q}^H&=\dfrac{1}{\pi}\int_0^\infty\left(f'(1-g^2)-gg'\sin2f+2f'g^2\sin^2f\right)dr\notag\\
    &=\dfrac{1}{\pi}\int_0^\infty\dfrac{d}{dr}\left(f-\dfrac{1}{2}g^2\sin2f\right)dr\notag\\
    &=\dfrac{1}{\pi}\left[f-\dfrac{1}{2}g^2\sin2f\right]^\infty_0.\label{top_charge_in_hedgehog}
\end{align}
It is straightforward to derive the equations which govern the critical points of (\ref{Energy_seq_hedgehog}). These are the coupled non-linear second-order $ODEs$ given by
\begin{align}
    2\kappa_0\left(r^2f''+2rf'-g^2\sin2f\right)-\kappa_2\left(gg''+\dfrac{g^2}{r^2}\left(1-g^2\right)\right)\sin2f\qquad&\label{E_a-EL-eq-f}\\
    +4\kappa_1\left(f'^2g^2\sin2f+\left(2f''g^2+4f'gg'-\dfrac{g^4}{r^2}\sin2f\right)\sin^2f\right)&=0,\notag\\
    g''+\dfrac{g}{r^2}\left(1-g^2\right)-4\kappa_0g\sin^2f-8\kappa_1\left(\dfrac{g^2}{r^2}\sin^2f+f'^2\right)g\sin^2f\qquad&\label{E_a-EL-eq-g}\\
    -\kappa_2\left(2f'^2g\cos2f+\left(f''g+2f'g'\right)\sin2f+2\left(g''+\dfrac{g}{r^2}\left(1-2g^2\right)\right)\sin^2f\right)&=0.\notag
\end{align}
\subsection{Skyrme-monopoles}
Recall that the corresponding Skyrme field given by the holonomy of a $(m,0)$-caloron, that is a BPS\footnote{We adopt the terminology `BPS monopole' for the remainder to distinguish monopoles on $\R^3$ and Skyrme-monopoles.} monopole, satisfies the boundary condition (\ref{Skyrme-monopole-bdy-cond}). There is hence no reason why we cannot consider boundary conditions like (\ref{Skyrme-monopole-bdy-cond}) for gauged skyrmions. A gauged skyrmion satisfying boundary conditions including the condition (\ref{Skyrme-monopole-bdy-cond}) will be called a \textbf{Skyrme-monopole} parameterised by $\nu\in(0,\pi]$.

At the moment, we are concerned with the spherically symmetric configurations. We hence consider the following boundary conditions for the profile functions within the spherically symmetric ansatz (\ref{hedgehog_ansatz}):
\begin{align}\label{Skyrme_monopole_BCs}
    \begin{array}{ll}
    f(0)=0,&g(0)=1,\\
    f(\infty)=\nu,&g(\infty)=0,
    \end{array}
\end{align}
for some constant $\nu\in(0,\pi]$. These conditions are chosen so that (\ref{Skyrme-monopole-bdy-cond}) holds, $U$ and $B$ are well-defined at $r=0$, and so that $D^BU,F^B\to0$ as $r\to\infty$ ensuring finite energy. Immediately, from (\ref{top_charge_in_hedgehog}), we see that such a Skyrme-monopole has topological charge $\mathcal{Q}^H=\nu/\pi$. We remark that boundary conditions of this type have previously been investigated for $SU(2)$ gauged Skyrme models in \cite{BrihayeHartmannTchhrakian2001MonopolesGaugeSky}.

There is a charge $m=1$ BPS monopole with spherical symmetry given by the monopole of Prasad and Sommerfield \cite{PrasadSommerfield1975}. The corresponding gauged Skyrme fields have the profile functions
\begin{align}
   f(r)&=\nu\coth(2\nu r)-\dfrac{1}{2r},\label{BPS-profile_f}\\
   g(r)&=\dfrac{2\nu r}{\sinh(2\nu r)},\label{BPS-profile_g}
\end{align}
and these are seen to satisfy the boundary conditions (\ref{Skyrme_monopole_BCs}). We may therefore compare gauged skyrmion configurations satisfying the field equations (\ref{E_a-EL-eq-f})-(\ref{E_a-EL-eq-g}), with the boundary conditions (\ref{Skyrme_monopole_BCs}), to this monopole.

Solving the equations (\ref{E_a-EL-eq-f})-(\ref{E_a-EL-eq-g}) explicitly is not so simple, so we shall approximate solutions numerically using a shooting algorithm. In order to do this, we must understand the limiting behaviour of the functions $f$ and $g$ near the boundaries. For $r<<1$, the linearisations of the field equations imply that there are constants $a,c>0$ such that
\begin{align}\begin{array}{cc}
    f(r)\sim aj_{1}\left(2\sqrt{c}r\right),&
    g(r)\sim 1-cr^2,
    \end{array}
\end{align}
where and $j_n(z)$ is the spherical Bessel function of the first kind
$$j_n(r)=(-r)^n\left(\frac{1}{r}\frac{d}{dr}\right)^n\frac{\sin r}{r}.$$
For $r$ large, the linearisation of the equation (\ref{E_a-EL-eq-f}) gives the large $r$ form of $f$ as
\begin{align}
    f(r)\sim\nu-\dfrac{b}{r},
\end{align}
for some constant $b\in\R$. This constant has a physical interpretation, namely the \textbf{scalar charge} of the Skyrme-monopole. To understand the asymptotics of $g$, we need to study equation (\ref{E_a-EL-eq-g}) with knowledge of the form of $f$. The linearisation is (up to order $r^{-1}$) given by
\begin{align}\label{g-equation-expansion-rinfty}
    g''-4\kappa_0\left(\dfrac{\sin^2\nu}{1-2\kappa_2\sin^2\nu}-\dfrac{\sin2\nu}{\left(1-2\kappa_2\sin^2\nu\right)^2}\dfrac{b}{r}\right)g+O(r^{-2})=0.
\end{align}
We expect an asymptotic form for $g$ of the form
\begin{align}
    g(r)=\sum_{j,k=0}^\infty e^{\chi_j r}r^{p_{jk}},
\end{align}
where $\chi_0>\chi_1>\dots$, and $p_{j0}>p_{j1}>\dots$ are real numbers for all $j\in\N$. In practice we only need to consider the leading term, so that $g(r)=e^{\chi r}r^p$. Plugging this into (\ref{g-equation-expansion-rinfty}), we may set the leading and sub-leading terms equal to $0$ and solve for $\chi$ and $p$. Choosing the decaying solution, we therefore have the asymptotic form for $g$ as
\begin{align}
    g(r)=d\exp\left(-2\sin\nu\sqrt{\dfrac{\kappa_0}{1-2\kappa_2\sin^2\nu}}r\right)r^{2b\cos\nu\sqrt{\kappa_0}\left(\dfrac{1}{1-2\kappa_2\sin^2\nu}\right)^{\frac{3}{2}}},\label{g-asym-SkyMon}
\end{align}
for some constant $d\in\R$ to be determined. It is important to note that the asymptotic formula (\ref{g-asym-SkyMon}) for $g$ near $r=\infty$ may only be applied for $\nu\neq\pi$, since when $\nu=\pi$, this formula does not decay unless $d=0$, which is not a reasonable choice for the asymptotics.

To find a suitable asymptotic form for $g$ when $\nu=\pi$, we shall need to include more terms in the expansion (\ref{g-equation-expansion-rinfty}). Consider now
$$f(r)\sim\pi-\dfrac{b}{r},$$
for some $b\in\R$. With this, we may expand (\ref{E_a-EL-eq-g}) with $g\sim 0$ to obtain
\begin{align}
    g''+\dfrac{4\kappa_2b^2}{r^3}g'+\left(1-4b^2\kappa_0\right)\dfrac{g}{r^2}+O(r^{-4})=0.\label{truncation_g_nu=pi}
\end{align}
Again we expect the asymptotic solution to take the form $g=de^{\chi r}r^p$. Substituting this ansatz into (\ref{truncation_g_nu=pi}), and setting the leading coefficients to $0$, we find a decaying solution given by
\begin{align}\label{g-asym-v=pi}
    g(r)\sim dr^{\frac{1}{2}\left(1-\sqrt{1+4\left(4b^2\kappa_0-1\right)}\right)},
\end{align}
again with $d\in\R$ constant. This is only decaying when 
\begin{align}\label{cut-off-condition-Sky-Mon-nu=pi}
    4b^2\kappa_0-1>0,
\end{align}
and if we do not have this condition, we expect no solutions to exist. This is in contrast to the case where $\nu\neq\pi$, where no such condition is required.
\subsubsection{Numerical results and comparison to monopoles}
In light of the results given in section \ref{section_inst-limit}, that as $\alpha\to\infty$, the gauged Skyrme energy $E_\alpha$ reduces to the ordinary Skyrme energy, it would be most interesting to study the behaviour at the other extreme, namely $\alpha$ small. The smallest value of $\alpha>-\frac{1}{2}$ for which we may explicitly calculate the coefficients in the energy is when $\alpha=0$, and so this is the case in which we shall study the general Skyrme-monopoles in detail. This is also a particularly interesting extreme case due to the behaviour of the topological bound as a function of $\alpha$, seen in figure \ref{fig.top_bound_plot}.

\begin{figure}[ht]
\centering
\includegraphics[width=0.8\textwidth]{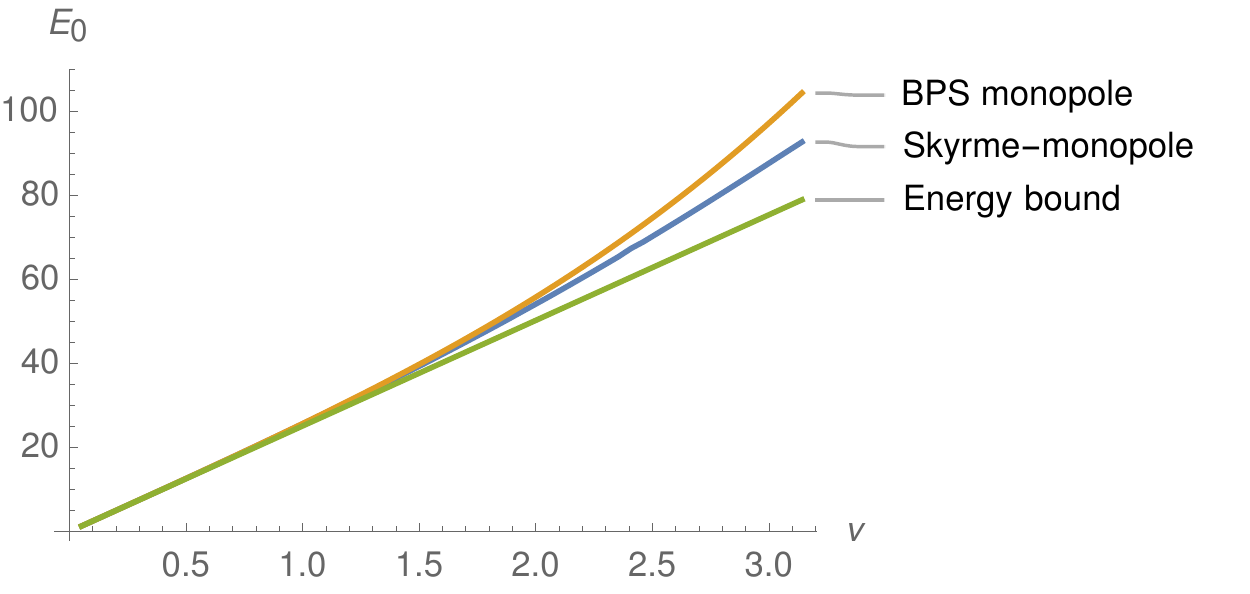}
\caption{The energies of the Skyrme-monopole (blue), monopole approximation (orange), and the topological absolute minimum (green), for $\nu\in(0,\pi]$.}
\label{fig.energy-comp-SkyMon}
\end{figure}

We find numerical solutions exist for all $\nu\in(0,\pi]$, and would like to compare the solutions to that of the BPS monopole. The energy data for both the numerical solutions and BPS monopole approximation are plotted in figure \ref{fig.energy-comp-SkyMon} along with the theoretical energy minimum given by the topological bound.
As can be seen, the approximation is really good, with less than a $1\%$ difference in the energies for $0<\nu<7\pi/15$, and negligible difference near $\nu=0$. As $\nu$ approaches $\pi$, the approximation is seen to be worse, but still within a reasonable degree of accuracy, with the percentage difference at $\nu=\pi$ only $12.6\%$.

\begin{figure}[ht]
\centering
\includegraphics[width=0.8\textwidth]{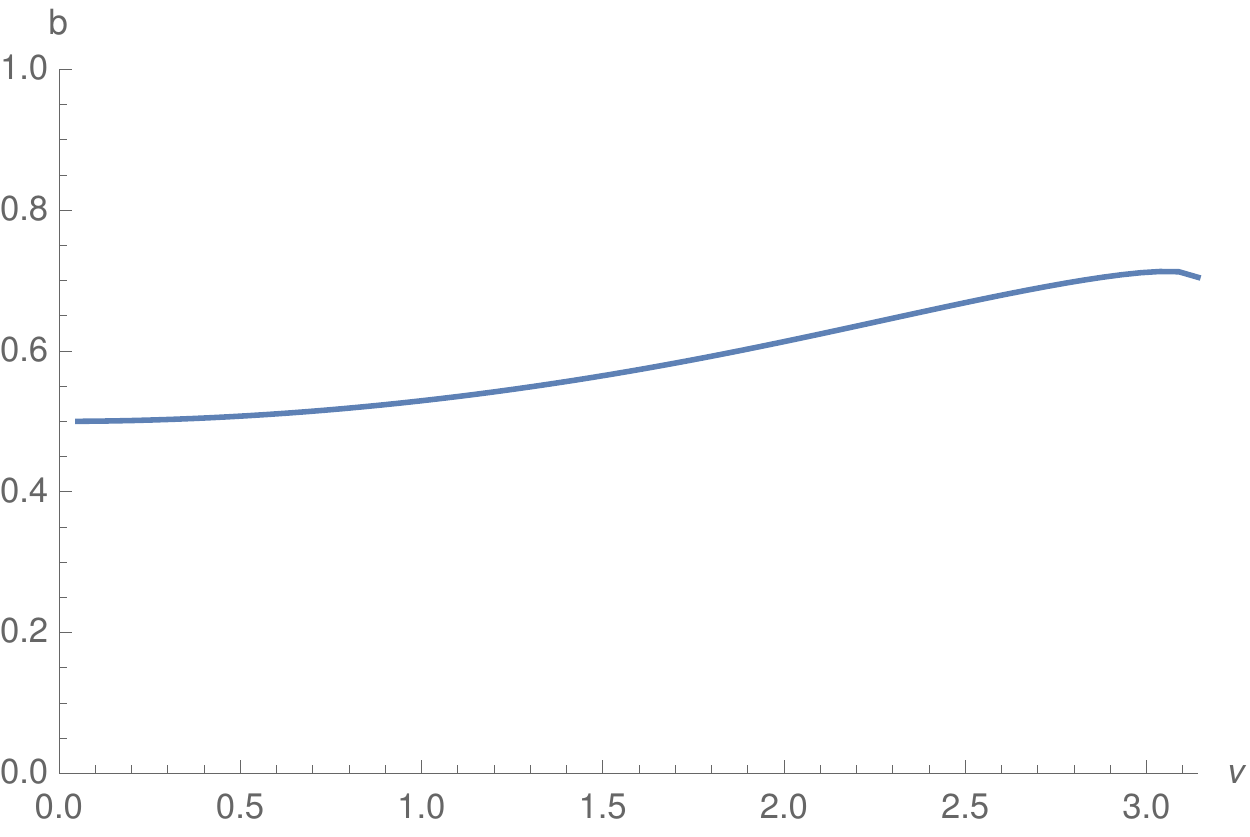}
\caption{The scalar charge of the Skyrme-monopoles for $\nu\in(0,\pi]$.}
\label{fig.B-plot}
\end{figure}

Another good measure of how the Skyrme-monopoles agree with the BPS monopoles is the scalar charge $b$. For the BPS monopole, this is a topological quantity, namely, it is exactly half the magnetic charge, i.e. $b=1/2$, as seen by the profile function (\ref{BPS-profile_f}). The value of $b$ for the Skyrme-monopole solutions is plotted in figure \ref{fig.B-plot}. As with the energy, this constant is seen to be close to that of the BPS monopole for $\nu\approx 0$, and deviates further away from that of the BPS monopole as $\nu\to\pi$.

\subsubsection{Varying \texorpdfstring{$\alpha$}{alpha}}
Having studied in some detail the Skyrme-monopole solutions for $\alpha=0$, it would be interesting to see what occurs as $\alpha$ varies. Doing this for general $\nu\in(0,\pi]$ is quite a laborious task, so instead we have considered three distinctly separated cases: $\nu=\pi$, $\nu=2\pi/3$, and $\nu=\pi/3$. The case $\nu=\pi$ is particularly interesting for two reasons. Firstly, the asymptotic behaviour for the profile function $g$ is explicitly different to that of $\nu\neq\pi$, and in particular the condition $4b^2\kappa_0(\alpha)-1>0$ must hold in order for solutions to exist. Secondly, the topological charge of such a Skyrme-monopole is $\mathcal{Q}=1$, which is the same as the configurations which we consider in section \ref{section-Skyrme_inst}. We may hence compare these Skyrme-monopoles with these other configurations, and this we shall also explore in section \ref{section-Skyrme_inst}.

\begin{figure}[ht]
\centering
\begin{subfigure}{.45\textwidth}
\centering
\includegraphics[width=0.8\linewidth]{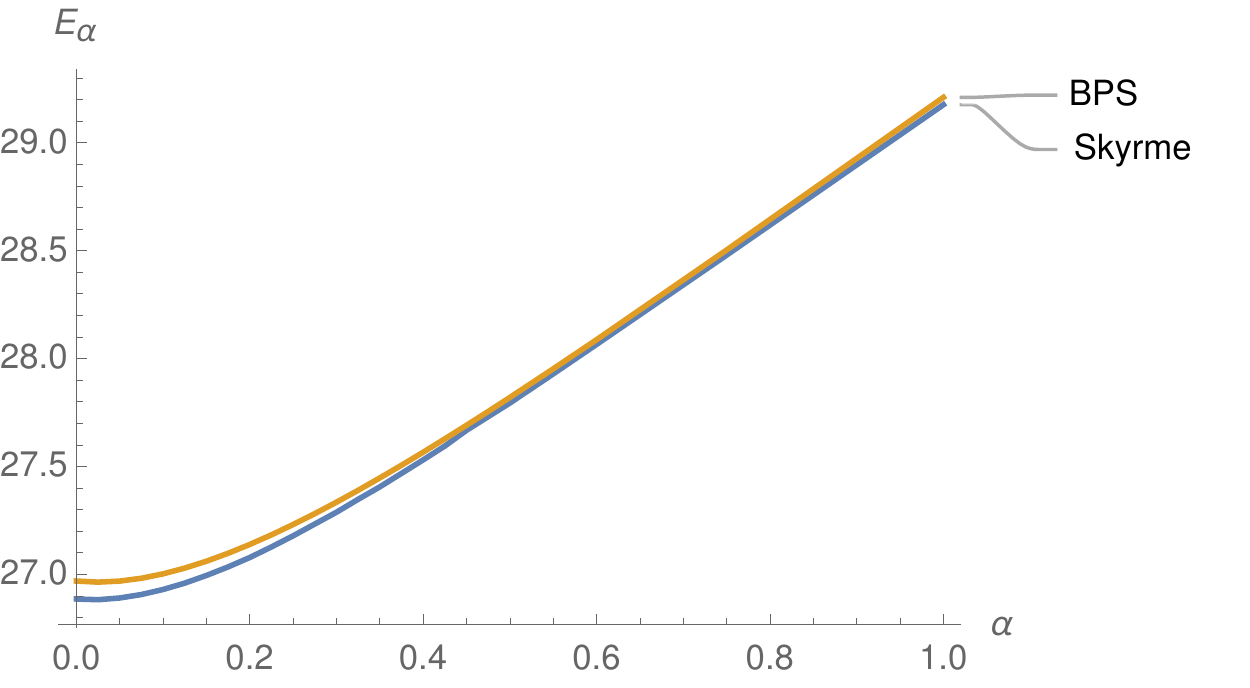}
\caption{$\nu=\frac{\pi}{3}$}
\label{subfig.energies-Sky-Mon-Pi/3}
\end{subfigure}
\begin{subfigure}{.45\textwidth}
\centering
\includegraphics[width=0.8\linewidth]{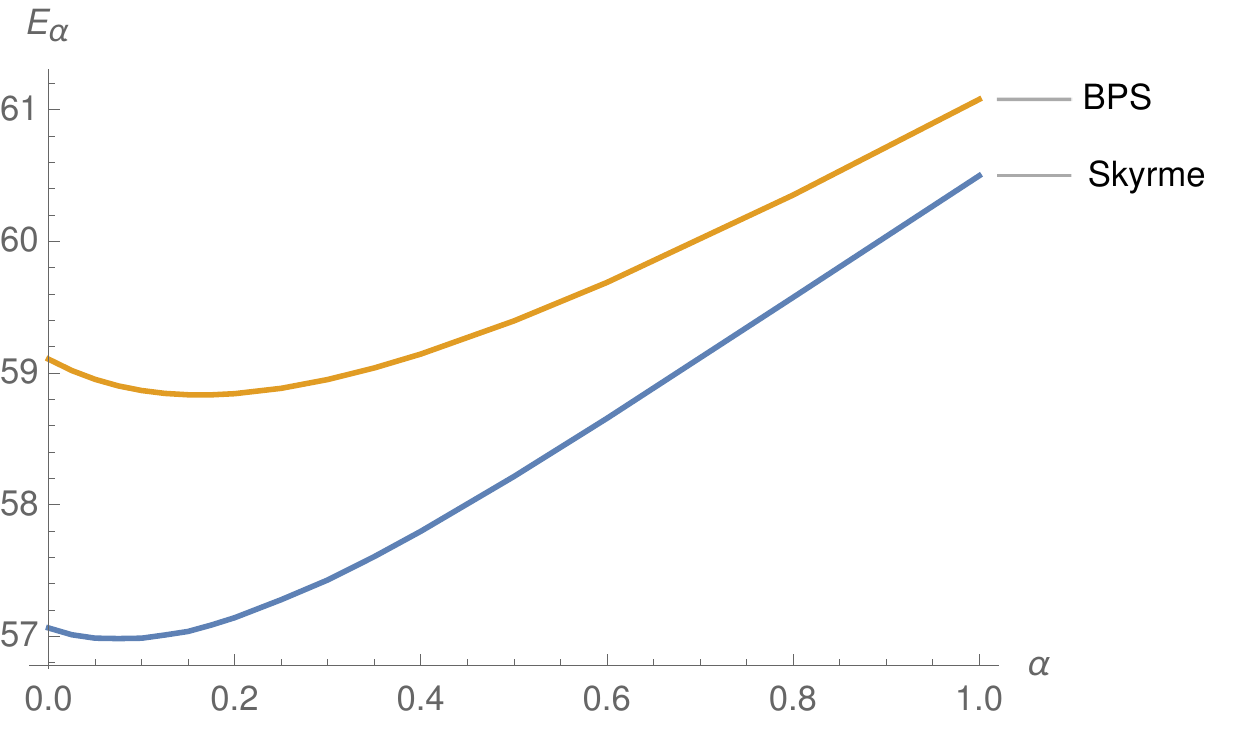}
\caption{$\nu=\frac{2\pi}{3}$}
\label{subfig.energies-Sky-Mon-2Pi/3}
\end{subfigure}
\begin{subfigure}{.45\textwidth}
\centering
\includegraphics[width=0.8\linewidth]{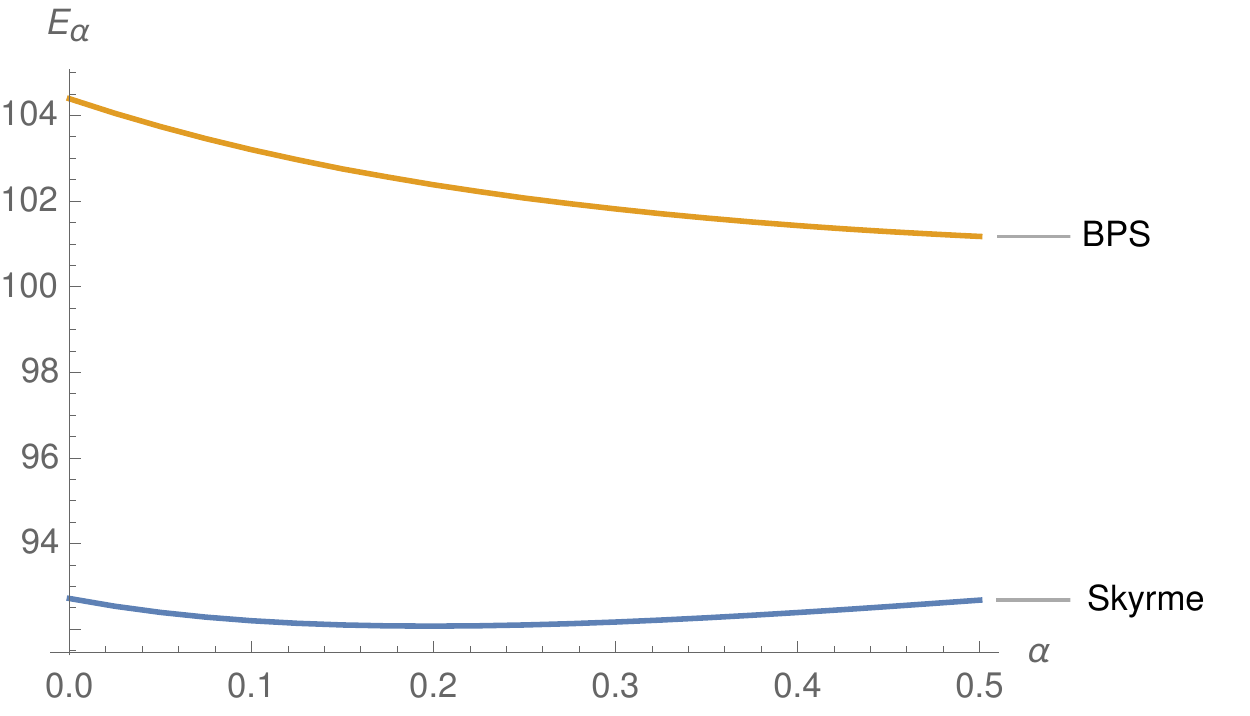}
\caption{$\nu=\pi$}
\label{subfig.energies-Sky-Mon-Pi}
\end{subfigure}
\caption{The phase diagrams for the energies of Skyrme and BPS monopoles with $\nu=\frac{\pi}{3},\frac{2\pi}{3}$, and $\pi$, for varying $\alpha\geq 0$.}
\label{fig.energies-Sky-Mon-varying}
\end{figure}

The phase diagrams of the energies $E_\alpha$ for these Skyrme-monopoles, compared to the BPS monopole energies, for varying $0\leq\alpha\leq1$, are plotted in figures \ref{subfig.energies-Sky-Mon-Pi/3}-\ref{subfig.energies-Sky-Mon-Pi}.
A similar observation occurs as with the detailed analysis of the case $\alpha=0$, namely, the BPS monopole approximation is significantly better for $\nu$ close to $0$ compared to $\nu$ close to $\pi$. It is also noticeable that as $\alpha$ increases, whilst the approximation improves in each case, the energies also increase. This seemingly monotone behaviour in the Skyrme-monopole energies is in contrast to the topological energy bound in figure \ref{fig.top_bound_plot}, which remains essentially constant for $\alpha>0$.

\begin{figure}[ht]
\centering
\includegraphics[width=0.8\textwidth]{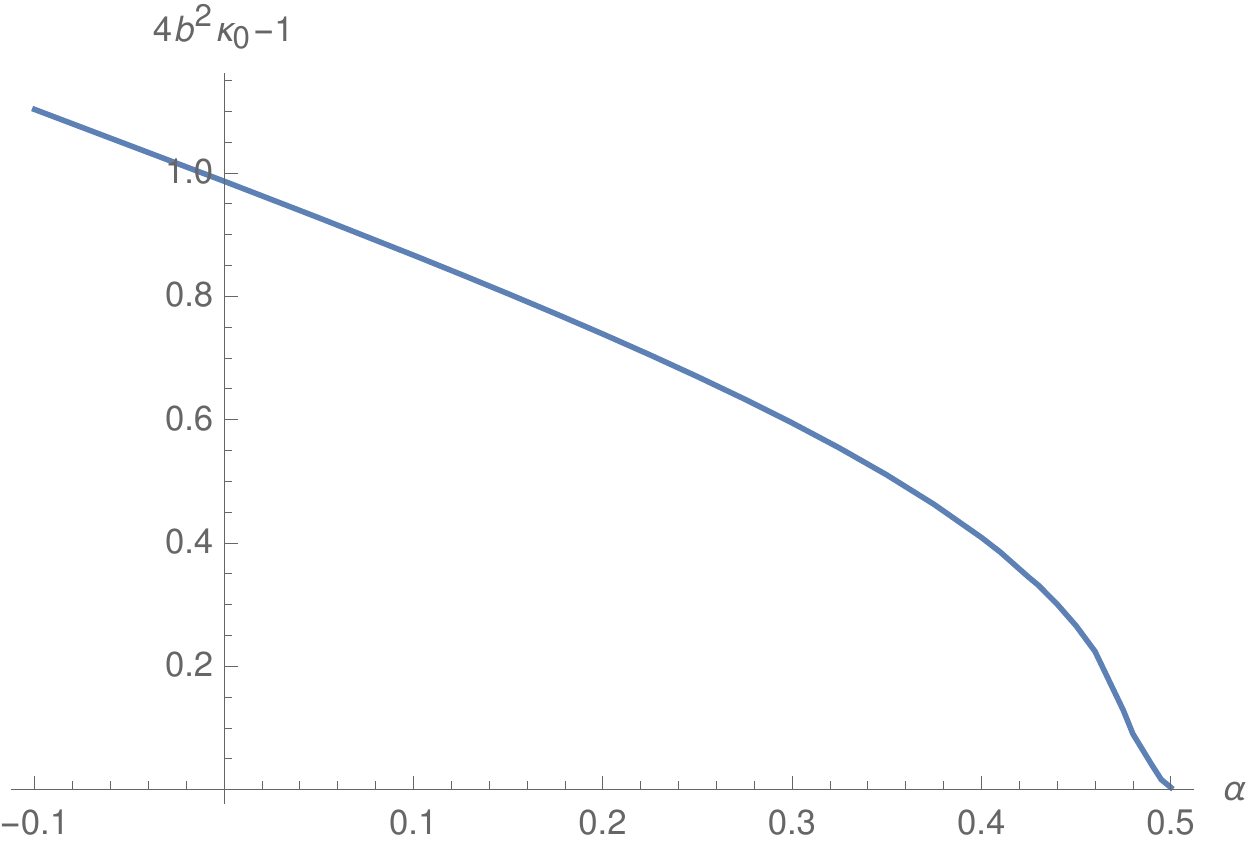}
\caption{The value of the `cut-off variable' $4b^2\kappa_0-1$ plotted as a function of $\alpha$.}
\label{fig.cutoff}
\end{figure}

Specifically in the case $\nu=\pi$, we find that for $-\frac{1}{2}<\alpha\leq\frac{1}{2}$,\footnote{We have only plotted for $\alpha\geq 0$ for the sake of clarity.} there are accurate numerical solutions. However, we were not able to generate such solutions for $\alpha>\frac{1}{2}$. In figure \ref{fig.cutoff}, we plot the quantity $4b^2\kappa_0-1$ for the $\nu=\pi$ Skyrme-monopoles, and it is observed that $4b^2\kappa_0(0.5)-1\approx0$,\footnote{The actual value we obtained numerically was $0.0036$ to two significant figures.} which may explain why our numerics did not behave well for $\alpha>0.5$. This aligns with the necessary condition (\ref{cut-off-condition-Sky-Mon-nu=pi}). 
Having said this, there is no reason to believe that the quantity $4b^2\kappa_0-1$ does not become positive again, and hence that additional Skyrme-monopoles with $\nu=\pi$ could exist, for some value of $\alpha>\frac{1}{2}$. This we are yet to investigate.

Rather intriguingly, for $\alpha=0$ we observe that $4b^2\kappa_0-1=1$ within numerical accuracy. From (\ref{g-asym-v=pi}), this means that for $\alpha=0$, $g$ behaves like $r^{-1/\varphi}$, where
$$\varphi=\dfrac{1+\sqrt{5}}{2}$$
is the golden ratio. It is unclear whether any meaning should be taken from this numerological observation, nevertheless, it is certainly rather curious.
\subsection{Skyrme-instantons}\label{section-Skyrme_inst}
In this section we consider Skyrme fields analogous to those constructed from the holonomy of a $(k,k)$-caloron. Recall that these satisfy the boundary condition (\ref{Skyrme-inst-bdy}):
\begin{align*}
    U\longrightarrow\begin{pmatrix}e^{-\imath\mu}&0\\
    0&e^{\imath\mu}
    \end{pmatrix},
\end{align*}
with $\mu\in[0,\pi]$, and have topological charge $\mathcal{Q}=k$. A gauged skyrmion satisfying boundary conditions including the condition (\ref{Skyrme-inst-bdy}) for any $\mu\in[0,\pi]$ will be called a \textbf{Skyrme-instanton} of degree $k$, where $k=\mathrm{deg}(U)$.

For the time-being, we are interested in the spherically symmetric examples. There is a one-parameter family of $(1,1)$-calorons which possess $O(3)$-symmetry. These are found within the family of Harrington-Shepard calorons \cite{HarringtonShepard1978periodic}, which are $(1,1)$ calorons with $\mu=0$. These calorons are often referred to as having \textit{trivial holonomy}. The components of the caloron gauge field are given explicitly by
 \begin{align}
     A_t(t,\vec{x})=\imath f(t,r)\dfrac{\vec{x}\cdot\vec{\sigma}}{r},\quad A_j(t,\vec{x})=\dfrac{\imath}{2}\left((g(t,r)-1)\dfrac{\epsilon_{jkl}x^k\sigma^l}{r^2}+h(t,r)\sigma^j\right),\label{Harr-shepp-1,1-gf}
\end{align}
where
\begin{align}
    f=-\dfrac{\bdy_r\phi}{2\phi},\quad g=1+\dfrac{r\bdy_r\phi}{\phi},\quad h=\frac{\bdy_t\phi}{\phi},
\end{align}
with $\phi:S^1\times\R^3\longrightarrow\R$ given by 
\begin{align}
     \phi=1+\dfrac{\lambda^2}{2r}\dfrac{\sinh(2\pi r)}{\cosh(2\pi r)-\cos(2\pi(t-\theta))}.
 \end{align}
This solution depends on two parameters, $\lambda>0$, which is interpreted as the `scale' of the caloron, and $\theta\in[-1/2,1/2]$ representing the caloron's location in $S^1$.

At the moment, (\ref{Harr-shepp-1,1-gf}) only has $SO(3)$-symmetry, due to the appearance of the function $h(t,r)$. In order to obtain $O(3)$-symmetry we need to fix the parameter $\theta$. Setting $\theta=\pm\frac{1}{2}$, or $\theta=0$ makes $h(-t,r)=-h(t,r)$, and then we have full spherical symmetry. In particular, in these cases $h(-1/2,r)=0$, so the constructed gauged Skyrme field is $O(3)$-symmetric. In these cases, the functions $f$ and $g$ take the forms\footnote{We have evaluated $g$ at $t=-\frac{1}{2}$ as this is the requirement to define the skyrmion gauge field $B$.}
\begin{align*}
    f_0(t,r)&=\dfrac{\lambda^2}{4r^2}\dfrac{\sinh(2\pi r)(\cosh(2\pi r)-\cos(2\pi t))-2\pi r\left(1-\cosh(2\pi r)\cos(2\pi t)\right)}{(\cosh(2\pi r)-\cos(2\pi t))(\cosh(2\pi r)-\cos(2\pi t)+\frac{\lambda^2}{2r}\sinh(2\pi r))},\\
    g_0(-\frac{1}{2},r)&=\dfrac{\pi \lambda^2+1+\cosh(2\pi r)}{\cosh(2\pi r)+1+\frac{\lambda^2}{2r}\sinh(2\pi r)},
\end{align*}
when $\theta=0$, and
\begin{align*}
    f_{\pm}(t,r)&=\dfrac{\lambda^2}{4r^2}\dfrac{\sinh(2\pi r)(\cosh(2\pi r)+\cos(2\pi t))-2\pi r\left(1+\cosh(2\pi r)\cos(2\pi t)\right)}{(\cosh(2\pi r)+\cos(2\pi t))(\cosh(2\pi r)+\cos(2\pi t)+\frac{\lambda^2}{2r}\sinh(2\pi r))},\\
    g_\pm(-\frac{1}{2},r)&=\dfrac{\cosh(2\pi r))-1-\pi\lambda^2}{\cosh(2\pi r)-1+\frac{\lambda^2}{2r}\sinh(2\pi r)},
\end{align*}
when $\theta=\pm\frac{1}{2}$. Note that since $\lambda\neq0$, we have $g_\pm\to -1$ as $r\to 0$, so the constructed Skyrme gauge field $B$ would have a singularity at $r=0$, as seen by the formula (\ref{hedgehog_ansatz}). This is problematic. However, no such singularity exists for the case $\theta=0$, since $g_0\to 1$ as $r\to0$, so we shall from now on only consider this case.

\begin{figure}[ht]
\centering
\includegraphics[width=0.8\textwidth]{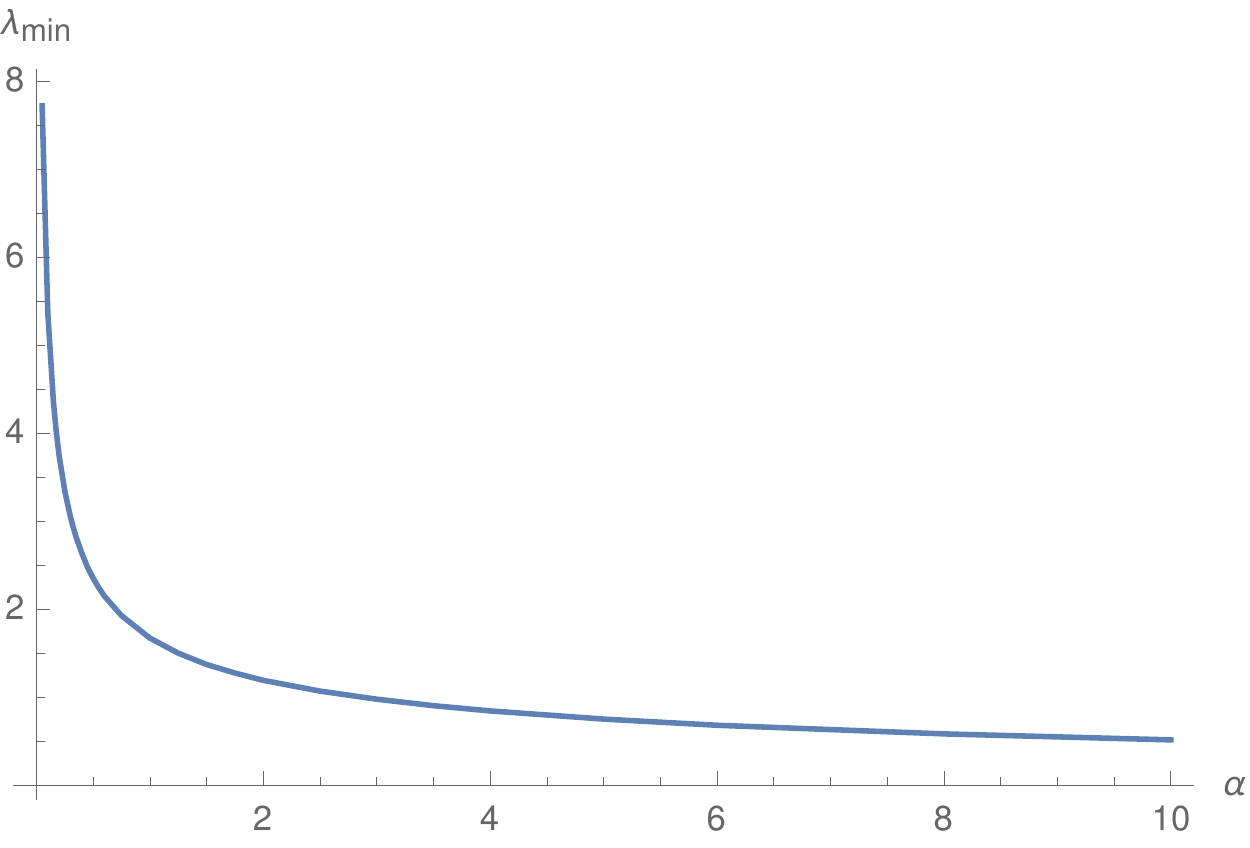}
\caption{The optimal value $\lambda_{\mathrm{min}}(\alpha)$ of the Harrington-Shepard scale parameter such that $E_\alpha$ is minimised.}
\label{fig.lambda-min}
\end{figure}

After a short calculation, we find that the resulting profile functions for the corresponding Skyrme fields are
\begin{align*}
    f_\lambda(r)&=\dfrac{\pi\sinh(2\pi r)-\frac{\lambda^2}{4 r^2}\left(\sinh(2\pi r)-2\pi r\cosh(2\pi r)\right)}{\sqrt{\left(\frac{\lambda^2}{2r}\sinh(2\pi r)+\cosh(2\pi r)-1\right)\left(\frac{\lambda^2}{2r}\sinh(2\pi r)+\cosh(2\pi r)+1\right)}}-\pi,\\
    g_\lambda(r)&=1+\dfrac{\lambda^2\pi-\frac{\lambda^2}{2r}\sinh(2\pi r)}{\frac{\lambda^2}{2r}\sinh(2\pi r)+1+\cosh(2\pi r)},
\end{align*}
where the scale $\lambda>0$ is the only remaining parameter. This parameter $\lambda$ may be optimised for each $\alpha$ so that $E_\alpha$ is minimised. We denote by $\lambda_{\mathrm{min}}$ this optimal value of $\lambda$ for each $\alpha>-\frac{1}{2}$. These optimal values are plotted for $\alpha>0$ in figure \ref{fig.lambda-min}.

\begin{figure}[ht]
\centering
\includegraphics[width=0.8\textwidth]{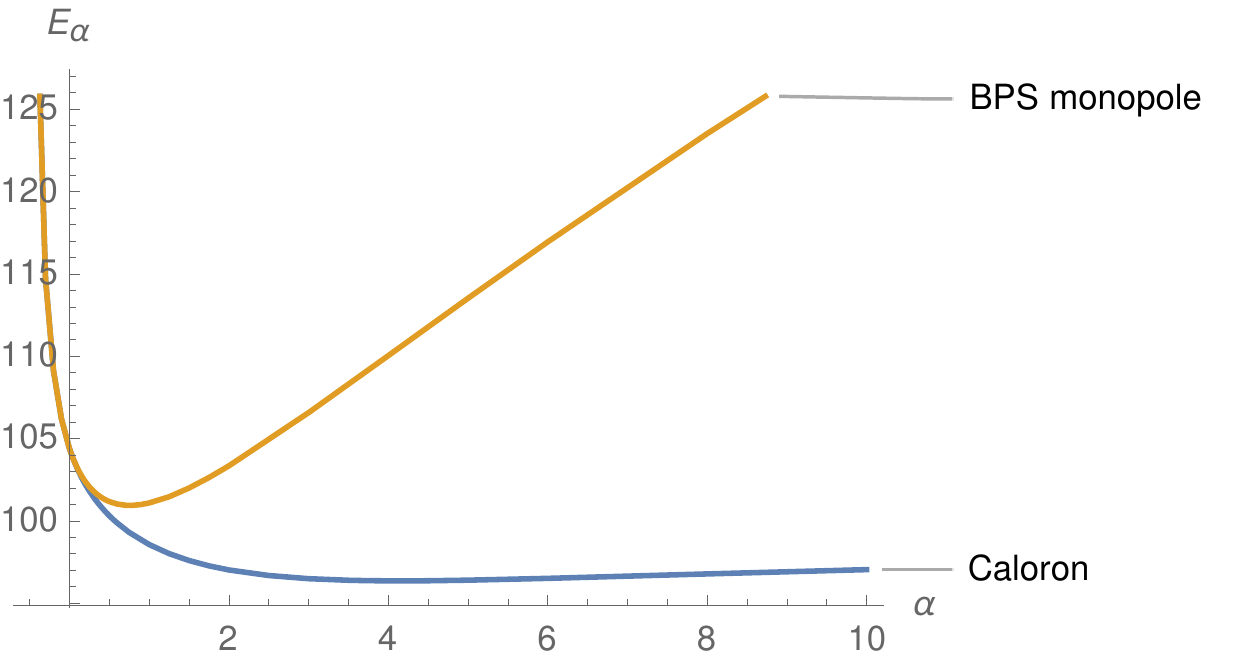}
\caption{The energies $E_\alpha$, for $-\frac{1}{2}<\alpha\leq10$, of the optimal Harrington-Shepard caloron and the $\nu=\pi$ BPS monopole.}
\label{fig.BPS-cal-energy}
\end{figure}

A noticeable property of this is that for $\alpha\sim0$, $\lambda_{\mathrm{min}}$ is very large. In fact, our numerics suggest that $\lambda_{\mathrm{min}}(0)=\infty$. Now, the functions $f_\lambda$ and $g_\lambda$ both have well-defined limits as $\lambda\to\infty$, given by
\begin{align}
    f_\infty(r)&=\pi\coth(2\pi r)-\dfrac{1}{2r}-\pi,\\
    g_\infty(r)&=\dfrac{2\pi r}{\sinh(2\pi r)}.
\end{align}
These are remarkably similar to the profile functions of the $\nu=\pi$ BPS monopole (\ref{BPS-profile_f})-(\ref{BPS-profile_g}), with the difference $f_\infty=f_{\mathrm{BPS}}-\pi$. This is not a coincidence. One can show that in the limit $\lambda\to\infty$ \cite{Harland2007,Rossi1979} that the Harrington-Sheppard $(1,1)$-caloron reduces to a $(0,1)$-caloron, which is equivalent, via a large gauge transformation called \textit{the rotation map} \cite{cork2018symmrot}, to the charge $1$ BPS monopole considered in the previous section. This observation, in light of figure \ref{fig.lambda-min}, suggests that the energy $E_\alpha$ prefers `monopole-like' boundary conditions near $\alpha=0$, and these become less preferred as $\alpha\to\infty$. This can also be seen by plotting the energy $E_\alpha(\lambda_{\mathrm{min}})$ against the energy $E_\alpha(\mathrm{BPS})$, which we do in figure \ref{fig.BPS-cal-energy}.

Having studied the behaviour of the caloron approximations to Skyrme-instantons, we would now like to see how this compares to the behaviour of the `true' Skyrme-instantons. We consider the following boundary conditions for the hedgehog profile functions:
\begin{align}\label{Skyrme_instanton_BCs}
    \begin{array}{ll}
    f(0)=-\pi,&g(0)=1,\\
    f(\infty)=0,&g(\infty)=1.
    \end{array}
\end{align}
These boundary conditions make (\ref{hedgehog_ansatz}) a Skyrme-instanton in the sense that the boundary condition (\ref{Skyrme-inst-bdy}) with $\mu=0$ holds. In particular, they are comparable to the Harrington-Shepard caloron, whose profile functions also obey these boundary conditions. From (\ref{top_charge_in_hedgehog}), the topological charge of such a Skyrme-instanton is $\mathcal{Q}^H=1$. We also remark that the boundary conditions (\ref{Skyrme_instanton_BCs}) are similar to those considered in \cite{ArthurTchrakian1996GaugedSky}.

In a similar way to the asymptotic analysis of the Skyrme-monopoles, we may linearise the field equations (\ref{E_a-EL-eq-f}) and (\ref{E_a-EL-eq-g}) to obtain formulae for the asymptotic behaviour of the Skyrme-instanton. We obtain
\begin{align}
    f_s(r)&=ar-\pi,\quad \text{for }r\sim 0,\label{Skyrme-inst-asym-f-0}\\
    f_l(r)&=-\dfrac{b}{r^2},\quad\text{for }r\sim\infty,
\end{align}
and
\begin{align}
    g_s(r)&=1-cr^2,\quad \text{for }r\sim 0,\\
    g_l(r)&=1-\dfrac{d}{r},\quad\text{for }r\sim\infty,\label{Skyrme-inst-asym-g-infty}
\end{align}
where the numbers $a,b,c,d\in\R$ may be determined using a Newton-Raphson shooting algorithm, analogously to the case of Skyrme-monopoles.

\subsubsection{Numerical results and comparison to calorons}
\begin{figure}[ht]
\centering
\includegraphics[width=0.8\textwidth]{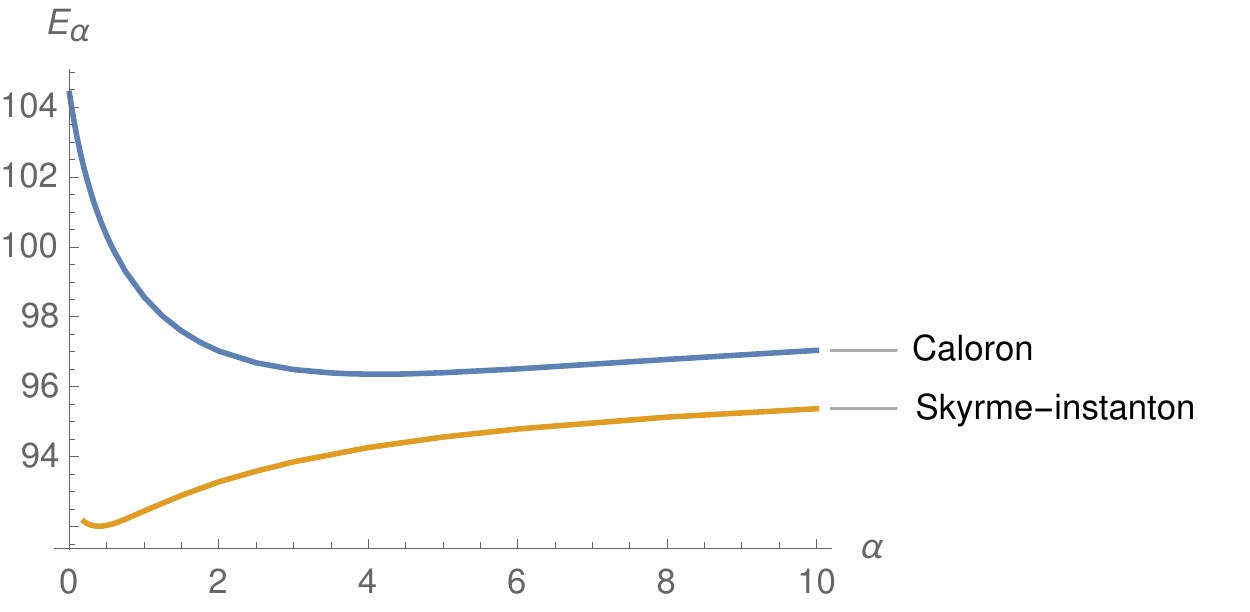}
\caption{The energies $E_\alpha$ for the optimal Harrington-Shepard caloron and the numerical Skyrme-instanton minimisers.}
\label{fig.Sky-cal-comp}
\end{figure}

For $\alpha>0$, and sufficiently not close to $0$, we find that there are numerical Skyrme-instanton solutions, and we plot their energies against the energies of the optimal caloron approximation in figure \ref{fig.Sky-cal-comp}, up to $\alpha=10$. It would appear from this plot that the caloron approximation gets better as $\alpha$ increases towards the weak coupling limit.
The behaviour near $\alpha=0$, the strong coupling limit, is interesting. In the case of Skyrme-monopoles with $\nu=\pi$, we found that there was a cut-off for which our numerics no longer returned valid solutions. For the Skyrme-instantons, our numerics reveal that near $\alpha=0$, the same absence of solutions occurs. However, in contrast to the case of the Skyrme-monopoles, here we do not have a reasonable hypothesis akin to (\ref{cut-off-condition-Sky-Mon-nu=pi}) which explains this. What we do have is the caloron approximation. This struggle to find solutions near $\alpha=0$ was actually predicted by the analysis of the Harrington-Shepard caloron, which suggested that monopole boundary conditions are preferred for $\alpha\sim0$. It would seem that this is the case for the actual solutions too.

\begin{figure}[ht]
\centering
\begin{subfigure}{.5\textwidth}
\centering
\includegraphics[width=0.8\linewidth]{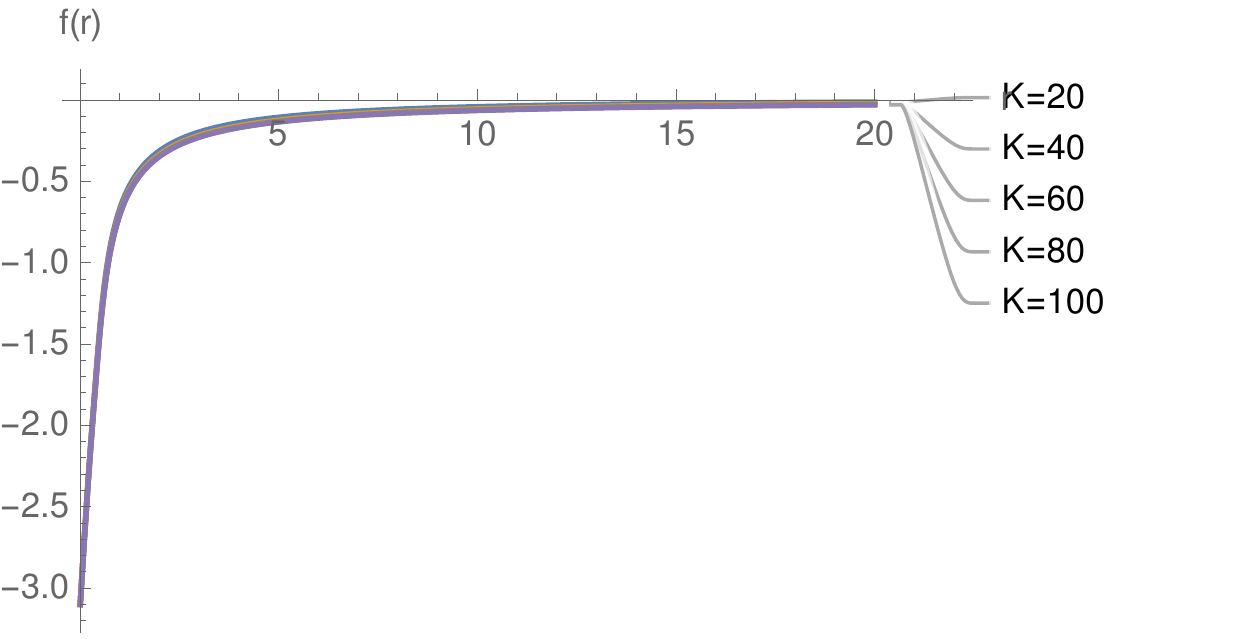}
\caption{$f$}
\label{subfig.E_0-f-functs-Sky-inst}
\end{subfigure}%
\begin{subfigure}{.5\textwidth}
\centering
\includegraphics[width=0.8\linewidth]{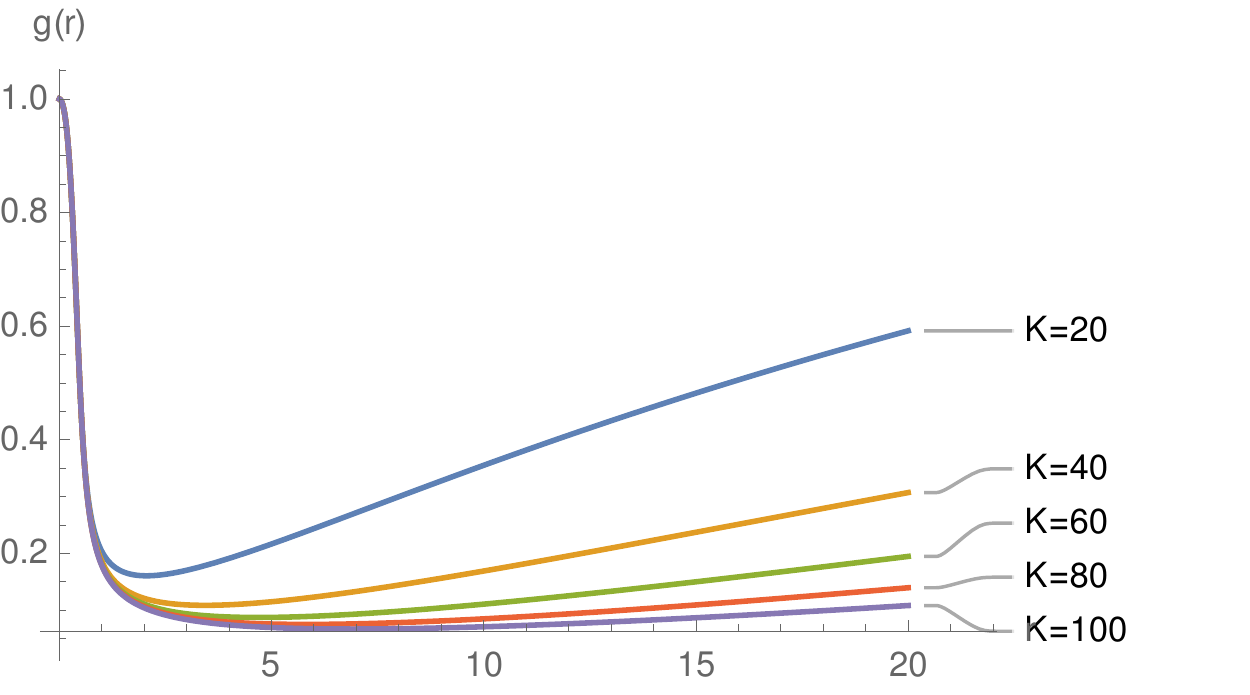}
\caption{$g$}
\label{subfig.E_0-g-functs-Sky-inst}
\end{subfigure}
\caption{The profile functions $f$ and $g$ for the Skyrme-instanton minimiser of $E_{0}$ on the finite intervals $[0.01,K]$, for $K=20,40,60,80,100$.}
\label{fig.E_0-f,g-functs-Sky-inst}
\end{figure}

As an example, we shall illustrate what occurs when we try shooting for numerical Skyrme-instanton solutions in the case $\alpha=0$. On each bounded interval $[\epsilon,K]$, for $\epsilon<<1$ and $K$ large, the Newton-Raphson algorithm for the constants $(a,b,c,d)$ appearing in the asymptotic formulae (\ref{Skyrme-inst-asym-f-0})-(\ref{Skyrme-inst-asym-g-infty}) converged. However, as the size of the interval $[\epsilon,K]$ was increased, the constant $d$ representing the large $r$ asymptotics of the profile function $g$, appeared to diverge. This can be seen in figure \ref{subfig.E_0-g-functs-Sky-inst}: as $K$ increases, the profile function $g$ for the gauge field $B$ does not converge to a function satisfying the boundary condition $g(\infty)=1$. Rather, it appears to become less and less localised, and more comparable to that of a Skyrme-monopole, satisfying $g\to0$ as $r\to\infty$. This is manifested in the constant $d=d(K)$ found in (\ref{Skyrme-inst-asym-g-infty}), which satisfies $d(20)\approx8.16$, and $d(100)\approx46.93$, suggesting that $g(\infty)$ would prefer to be $0$. In contrast, the profile function $f$ for the Skyrme field $U$ does appear to converge (see figure \ref{subfig.E_0-f-functs-Sky-inst}), which is expected since the boundary conditions are the same as the Skyrme-monopole under the replacement $f\mapsto f-\pi$, which is a symmetry of the field equations.

The energy calculated for this Skyrme-instanton is $E\approx92.74$, which is extremely close to that of the $\nu=\pi$ Skyrme-monopole, which has energy $E\approx92.72$. We suspect that if we were to continue for $K>100$, then the energy of the Skyrme-instanton will lower, converging on the energy of the Skyrme-monopole.

A similar pattern in the numerics is observed for all $0\leq\alpha<0.1953$, that is, the algorithm did not converge. One hypothesis as to why this is the case for the small values of $\alpha\neq0$ is that the Skyrme-instanton actually does exist, but it is extremely large, and to construct it would require considering values of $K$ which are far greater than $100$, where we usually stopped the process. This is evidenced by considering the behaviour of the optimal Harrington-Shepard caloron profile functions, for which $g_\lambda$, for $\lambda$ large, does not get near to $1$ until $r$ is very large. Another idea is that our numerical algorithm is not robust enough to find all of the solutions. One alternative method is to use \textit{pseudo-arclength continuation} alongside our usual shooting algorithm, in which we would also vary $\alpha$, changing the shooting map to a function $F:\R^5\longrightarrow\R^4$, $F=F(\alpha,a,b,c,d)$. This method has been used for similar purposes, namely to pick out seemingly absent solutions to field equations which depend on a parameter, for example in \cite{FloodSpeight2018chern}. Of course, it is also possible that the algorithm did not converge because no solution with those boundary conditions exists.

\subsubsection{Comparison to Skyrme-monopoles}
The main conclusion of studying the spherically-symmetric Skyrme-instantons and Skyrme-monopoles is that the energy (\ref{Energy_seq_hedgehog}) appears to favour certain boundary conditions as $\alpha$ varies. To test this idea further, we will make another comparison. The Skyrme-monopoles with $\nu=\pi$, and the Skyrme-instantons, both have topological charge $\mathcal{Q}^H=1$, so it is reasonable to compare them as solitons. In fact, we have observed that when $\alpha\approx0$, these configurations may even be the same (up to a large gauge transformation), in analogy with the $\lambda\to\infty$ limit of the Harrington-Shepard caloron.

\begin{figure}[ht]
\centering
\includegraphics[width=0.8\textwidth]{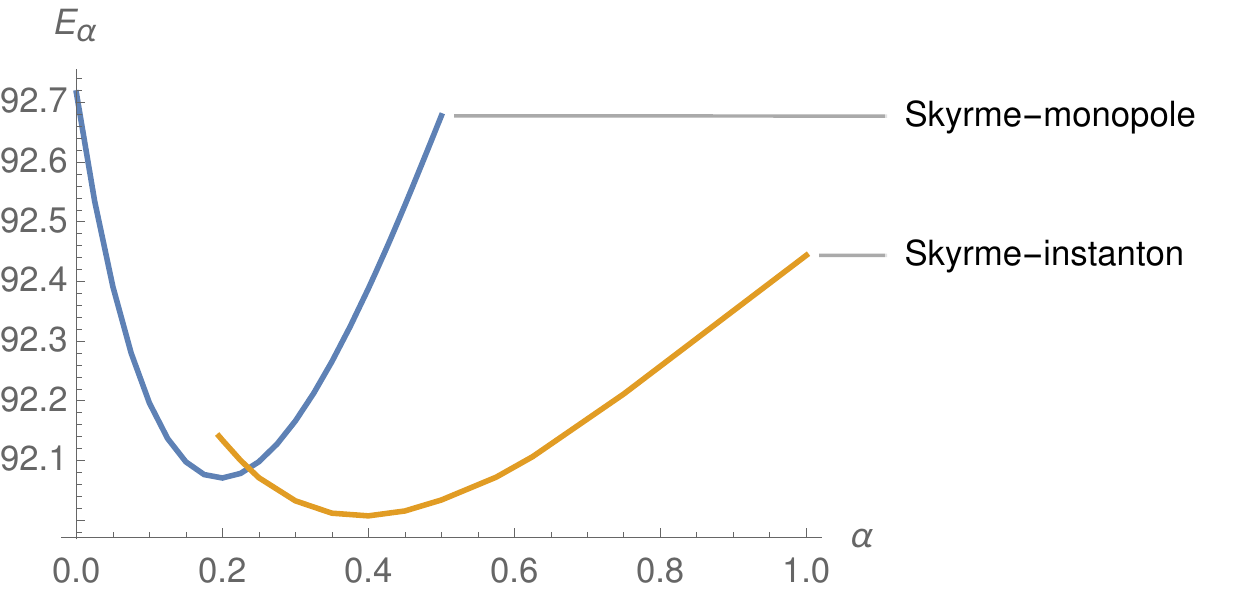}
\caption{A phase diagram of the energies of Skyrme-monopoles and Skyrme-instantons for varying $\alpha\in[0,1]$.}
\label{fig.phase-diagram-gauged-skyrmions}
\end{figure}

Consider the phase diagram in figure \ref{fig.phase-diagram-gauged-skyrmions}. There we have plotted the value of the energy $E_\alpha$ for both the numerical Skyrme-monopoles and Skyrme-instantons for $\alpha\in[0,1]$. Clearly, for $\alpha\approx0$, the Skyrme-monopole boundary conditions dominate since there are no Skyrme-instanton solutions. Extrapolating the curve for the Skyrme-instantons in such a way that the two curves meet at $\alpha=0$, it is easy to convince oneself that this is also true with regards to minimising the energy. Likewise, the Skyrme-instanton solutions dominate away from $\alpha=0$.

\section{Approximating skyrmions with gauged skyrmions}\label{sec.gauged-sky-to-sky}
An energy of the form (\ref{gauged_Skyrme_energy}) describing an $SU(2)$ gauged Skyrme model is a functional $E(U,B)$ of fields $(U,B)$, where $U:\R^3\longrightarrow SU(2)$, and $B$ is a connection $1$-form on $\R^3$. This naturally induces an ordinary Skyrme model in the case that $B=0$, with energy
\begin{align}
    E^o=&\int\left(\lambda_0|U^{-1}dU|^2+\lambda_1|U^{-1}dU\wedge U^{-1}dU|^2\right)d^3x.\label{Skyrme-energy-B=0}
\end{align}
This functional describes a bona-fide $SU(2)$ Skyrme model, whose critical points satisfy the Skyrme field equation (\ref{Skyrme-field-equation}) with $\lambda_p=c_p$. However, unlike the gauged model, this induced model is not invariant under gauge transformations $G:\R^3\longrightarrow SU(2)$, since here we have
\begin{align*}
    U^{-1}dU\mapsto G\left(U^{-1}dU+U^{-1}L_GU-L_G\right)G^{-1},
\end{align*}
where $L_G=G^{-1}dG$. Letting $\mathcal{L}=U^{-1}(dU+[L_G,U])$, the Skyrme energy (\ref{Skyrme-energy-B=0}) thus transforms as $E^o(U)\mapsto E(G,U)$, where
\begin{align}
    E(G,U)=\int\left(\lambda_0|\mathcal{L}|^2+\lambda_1|\mathcal{L}\wedge\mathcal{L}|^2\right)d^3x.\label{Skyrme-energy_GT}
\end{align}
In each gauge equivalence class of gauged skyrmions $(U,B)$ (that is, critical points of (\ref{gauged_Skyrme_energy})), it is not unreasonable to ask whether there is a representative $G\cdot(U,B)$ such that $GUG^{-1}$ approximates a critical point of (\ref{Skyrme-energy-B=0}). This is equivalent to saying $(G,U)$ approximates a critical point of (\ref{Skyrme-energy_GT}). Such a pair $(G,U)$ must satisfy the asymptotic boundary conditions $G,U\to\mathbb{1}$ as $|\vec{x}|\to\infty$, in line with the usual boundary conditions imposed on the Skyrme field.

The important variable that needs to be optimised here is the choice of gauge. Varying $E(G,U)$ with respect to $G$ gives the equation
\begin{align}\label{Gauge-fixing-minimisation-cond}
    \sum_{i,j}\bdy_i\left(G\left[\lambda_0\mathcal{L}_i+\lambda_1\left[\mathcal{L}_j,[\mathcal{L}_i,\mathcal{L}_j]\right],U\right]U^{-1}G^{-1}\right)=0,
\end{align}
which is a second order partial differential equation for $G$. So for any gauged skyrmion $(U,B)$, that is a critical point of (\ref{gauged_Skyrme_energy}), the representative $U'$ which minimises (\ref{Skyrme-energy-B=0}) is hence given by $U'=GUG^{-1}$, where $(G,U)$ solves (\ref{Gauge-fixing-minimisation-cond}).

Imposing a symmetric form on the Skyrme field $U$ simplifies this condition. For example, when $(U,B)$ is spherically symmetric, i.e. of the form in (\ref{hedgehog_ansatz}), the gauge transformations which preserve this are those of the same `hedgehog' form:
\begin{align}
    G(\vec{x})=\exp\left(\imath\mu(r)\dfrac{\vec{x}\cdot{\vec{\sigma}}}{r}\right),
\end{align}
for some function $\mu:(0,\infty)\longrightarrow\R$. In this scenario, $G$ acts trivially on $U$, that is $GUG^{-1}=U$, and so the condition (\ref{Gauge-fixing-minimisation-cond}) is obsolete. In other words, the energy (\ref{Skyrme-energy-B=0}) is invariant under gauge transformations of spherically-symmetric gauged skyrmions.

With this in mind, we may automatically compare the minimisers of (\ref{Skyrme-energy-B=0}), to the Skyrme-instantons and Skyrme-monopoles found in the previous section, without having to consider a preferred choice of gauge. We set $\lambda_0=\kappa_0$, and $\lambda_1=\frac{\kappa_1}{2}$ so that (\ref{Skyrme-energy-B=0}) aligns with (\ref{family-gauged-energies}). Within the hedgehog ansatz, the field equation for (\ref{Skyrme-energy-B=0}) reduces to the ODE
\begin{align}
    \left(\kappa_0r^2+4\kappa_1\sin^2f\right)f''+2\kappa_0rf'+\sin2f\left(2\kappa_1{f'}^2-\kappa_0-2\kappa_1\dfrac{\sin^2f}{r^2}\right)&=0.\label{E-ordinary-field-hedgehog}
\end{align}
The usual boundary conditions imposed for a spherically-symmetric skyrmion are \cite{MantonSutcliffe2004} $f(0)=\pi$ and $f(\infty)=0$. We shall instead consider the boundary conditions $f(0)=-\pi$ and $f(\infty)=0$ so that by the formula (\ref{top_charge_in_hedgehog}), the skyrmion $U=\exp\left(\imath f(r)\frac{\vec{x}\cdot\vec{\sigma}}{r}\right)$ satisfying (\ref{E-ordinary-field-hedgehog}) has topological charge $\mathcal{Q}^H=1$. This boundary condition is also comparable to the Skyrme fields of the Skyrme-instantons, and the Skyrme-monopoles with $\nu=\pi$, from the previous section.\footnote{We remark that $f\mapsto f+n\pi$ is a symmetry of the energy and topological charge for all $n\in\Z$, so the boundary condition $f(0)=0$, and $f(\infty)=\pi$ for the Skyrme-monopoles is essentially equivalent to this.} It is worthwhile noting that even though equation (\ref{E-ordinary-field-hedgehog}) appears to depend on the couplings $\kappa_0,\kappa_1$, that is, the parameter $\alpha$, this dependence is only the case up to a re-scaling of length and energy units. It follows that all solutions of (\ref{E-ordinary-field-hedgehog}) with these boundary conditions are the same up to this re-scaling, and will hence all be called \textit{the} spherically-symmetric skyrmion. Linearising (\ref{E-ordinary-field-hedgehog}) with these boundary conditions gives a Cauchy-Euler type equation, and we obtain the asymptotic formulae
\begin{align}
    f_s(r)=ar-\pi,\quad r\sim0,\\
    f_l(r)=-\dfrac{b}{r^2},\quad r\sim\infty,
\end{align}
for some $a,b\in\R^+$ to be determined numerically for each $\alpha>-\frac{1}{2}$.

We are particularly interested in comparing this ordinary skyrmion to the \textit{gauged skyrmions} considered in the previous sections, that is, the Skyrme-monopoles and Skyrme-instantons. The most enlightening comparison in this situation is how much the profile functions agree (or disagree) with each other. To measure this, we will calculate $\Xi=\max_{r}|f-f_{\mathrm{Sky}}|$, where $f$ is the profile function of the gauged skyrmion Skyrme fields, to be varied over the different types, and $f_{\mathrm{Sky}}$ is the ordinary skyrmion (scaled accordingly to solve (\ref{E-ordinary-field-hedgehog}) for the correct value of $\alpha$). We will also use the same measure when we come to compare the ordinary skyrmion to the optimal Harrington-Shepard calorons, and the BPS monopoles, in the next section. Of course, for both types of monopoles, we consider instead $f-\pi$.

\begin{figure}[ht]
\centering
\begin{subfigure}{.49\textwidth}
\centering
\includegraphics[width=0.8\linewidth]{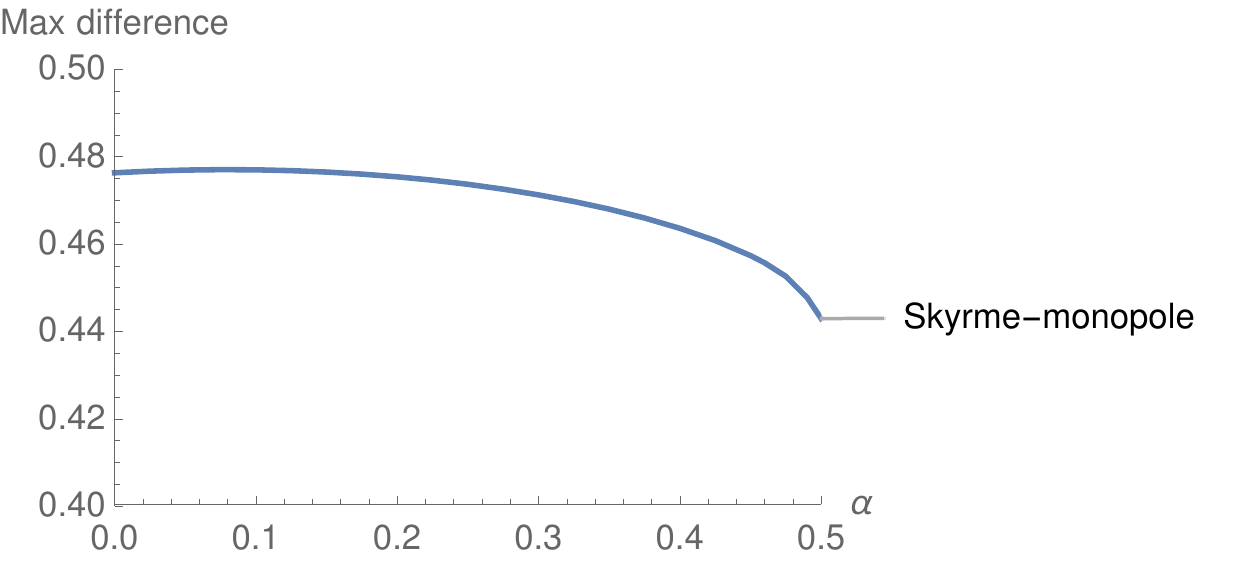}
\caption{ }
\label{subfig.Sky-mon-compare-ordinary}
\end{subfigure}
\begin{subfigure}{.49\textwidth}
\centering
\includegraphics[width=0.8\linewidth]{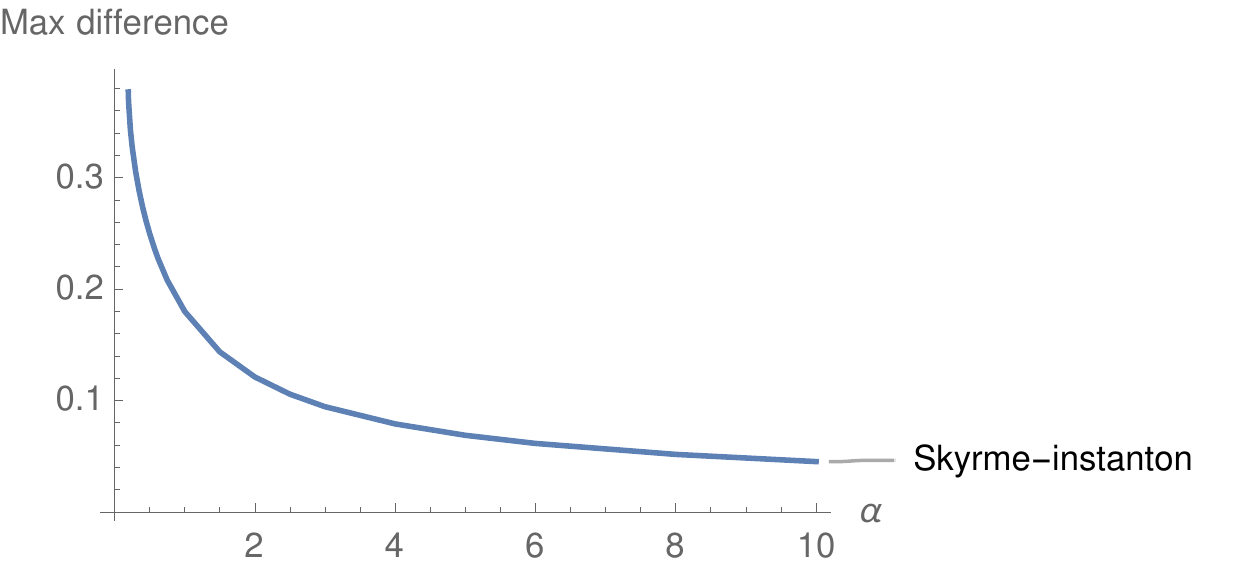}
\caption{ }
\label{subfig.Sky-inst-compare-ordinary}
\end{subfigure}
\caption{The maximum difference between the Skyrme-monopole/Skyrme-instanton Skyrme fields, and the ordinary Skyrme field, as a function of $\alpha$.}
\label{fig.Sky-mon-inst-compare-ordinary}
\end{figure}

In figures \ref{subfig.Sky-mon-compare-ordinary}-\ref{subfig.Sky-inst-compare-ordinary}, we plot this measure of difference between the ordinary skyrmion, against the Skyrme fields of the Skyrme-monopoles (with $\nu=\pi$) and Skyrme-instantons respectively.

In the case of the Skyrme-monopoles, the difference is seen to be the greatest of all configuration types studied, but still below $0.48$ for all examples, which is $\approx15\%$ of the maximum absolute value of the skyrmion's profile function ($|f(0)|=\pi$). As $\alpha$ varies, this measure of deviation is relatively constant, remaining between $0.45$ and $0.48$. Contrary to this, the difference between the Skyrme-instantons, and the ordinary skyrmions starts similarly to the Skyrme-monopoles, and then becomes almost negligible as $\alpha$ increases. This is in fact very much expected -- as a result of the discussion in section \ref{section_inst-limit}, we know that in the instanton/weak coupling limit $\alpha\to\infty$, the family of spherically-symmetric Skyrme-instantons that we have described will converge in some way to the ordinary spherically-symmetric skyrmion.
\subsection{Approximating skyrmions with calorons and monopoles}
To finish our discussions on calorons, gauged skyrmions, and skyrmions, we shall make one last set of comparisons. Part of the motivation for studying this topic was to see how well calorons, and in particular monopoles, approximate ordinary skyrmions. This has, as already mentioned, been investigated in part in \cite{EskolaKajantie1989skycal,NowakZahed1989skycal}, but without the knowledge of the family of gauged Skyrme models (\ref{family-gauged-energies}), and the intermediate relationship between calorons and gauged skyrmions. Furthermore, as far as we know, no such comparison between BPS monopoles and skyrmions has ever been made.

\begin{figure}[ht]
\centering
\includegraphics[width=0.8\textwidth]{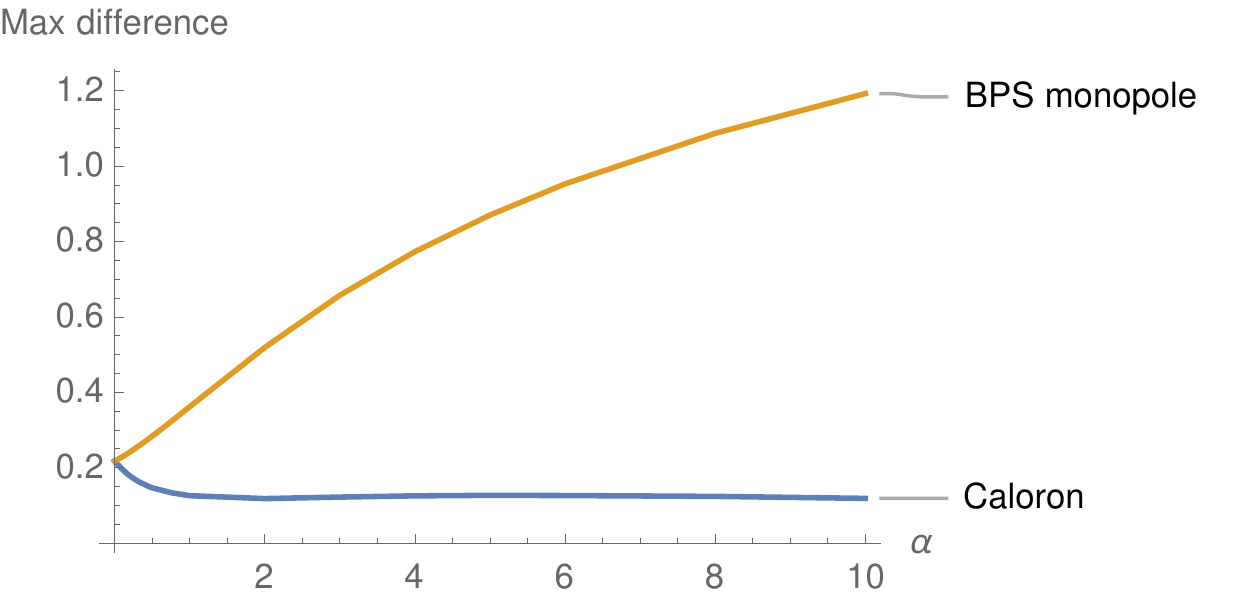}
\caption{The maximum difference between the Skyrme field profile function for the charge $1$ BPS monopole, and optimum Harrington-Shepard caloron, as a comparison with the ordinary spherically-symmetric skyrmion.}
\label{fig.BPS-Cal-compare-ordinary}
\end{figure}

In the same way as with the gauged skyrmions, we compare the optimal Harrington-Shepard calorons, and BPS monopoles with $\nu=\pi$, by measuring the maximum difference between the profile functions. The results for varying $0\leq\alpha\leq10$ are plotted in figure \ref{fig.BPS-Cal-compare-ordinary}.
In light of the observations in section \ref{section-Skyrme_inst}, it is unsurprising that the strength of the BPS monopole approximation diminishes as $\alpha$ increases, with the difference between the profile functions growing fairly rapidly. On the other hand, the Harrington-Shepard caloron approximation improves as $\alpha$ increases, but of course, this is expected as the caloron appeared to better approximate the Skyrme-instantons in this way, as seen in figure \ref{fig.Sky-cal-comp}. These two plots meet at $\alpha=0$, and importantly, the difference between them and the ordinary Skyrme field is small, at approximately $0.25$, which is only $8\%$ of the maximum absolute value of the skyrmion's profile function.\footnote{This is actually not the minimum value found. A smaller value of the max difference may be obtained at $\alpha\sim-0.2$, namely a difference of $0.21$.} The conclusion of this brief analysis is that calorons appear to be good approximations of skyrmions at all length and energy scales (for optimal choices of the parameter $\lambda$), and crucially, monopoles are a good approximation when the length and energy scales are those which align with the strong coupling $\alpha\approx0$.
\section{Summary and open problems}
By utilising the Atiyah-Manton-Sutcliffe methods for constructing approximate Skyrme fields from instantons \cite{AtiyahManton1989,sutcliffe2010skyrmions}, we have shown how to similarly construct approximate \textit{gauged Skyrme fields} from periodic instantons, also known as calorons. One nice property of this construction is that it considers an expansion of the caloron in terms of the ultra-spherical functions, which leads to a one-parameter family of gauged Skyrme models (\ref{family-gauged-energies}), parameterised by the ultra-spherical parameter $\alpha>-\frac{1}{2}$. In particular, this family, with any number of vector mesons included, reduces to the corresponding Sutcliffe model \cite{sutcliffe2010skyrmions} in a limit where $\alpha\to\infty$.

We have studied the relationships between calorons and gauged skyrmions in the case of spherically symmetric examples. The main conclusion is that the model appears to interpolate between different favoured boundary conditions as $\alpha$ varies, with `monopole-like' boundary conditions preferred as $\alpha\to0$, and `instanton-like' conditions preferred as $\alpha\to\infty$. This is rather interesting given the similar interpolation between monopoles and instantons that calorons exhibit. We have also studied how monopoles and calorons can be in some cases seen to be reasonable approximations to ordinary skyrmions. This is a small step of progress in further understanding the links between monopoles and skyrmions.

There is of course a lot of work still to be done here. Crucially, we have only considered the most basic examples, and a lot more is likely to be revealed by looking at less symmetric field configurations. The next most simple example would be to consider axially symmetric examples. The famous $(1,1)$-calorons with non-trivial holonomy \cite{KraanVanBaal1998,LeeLu1998}, contain a family of calorons with precisely this symmetry. Like with the Harrington-Shepard calorons, the holonomies of these calorons generate (less trivial) examples of Skyrme-instantons of charge $1$, with a scale parameter $\lambda>0$ which can be optimised. Since the Skyrme-instanton boundary condition (\ref{Skyrme-inst-bdy}) for non-zero $\mu$ breaks the gauge symmetry to $U(1)$, there may be a relationship with the $U(1)$-gauged skyrmions found in \cite{PietteTchrakian2000static}, which also exhibit an axial symmetry, in addition to a non-zero dipole moment, which is similar to the interpretation of $(1,1)$-calorons as two oppositely charged magnetic monopoles \cite{kraanVanBaal1998monopoleconsts}. Another axially symmetric caloron appears explicitly in \cite{KatoNakamulaTakesue2018magnetically}. This caloron is possibly more interesting as it has a mixture of instanton and magnetic charge -- it is a $(2,1)$-caloron. Moving on from axial symmetry, Ward presents in \cite{Ward2004} examples of $(k,k)$-calorons, with trivial holonomy, in the cases $k=2,3,$ and $4$, which exhibit platonic symmetries. In all of these cases, the monopole and instanton limits have been studied \cite{Harland2007,KraanVanBaal1998,LeeLu1998,Ward2004}. It would be very interesting to see whether there exist corresponding symmetric gauged skyrmions, how well these calorons approximate them, what their behaviour is with respect to the parameter $\alpha$, and of course, how they relate to the ordinary skyrmions with the same symmetries. In particular, it would be good to see if this comparison evinces the apparent correlation between the symmetries of certain monopoles and skyrmions \cite{HoughtonMantonSutcliffe1998rational}. It would also be interesting to study the more obscure symmetric examples of calorons in the context of skyrmions, for example the crossed solutions and oscillating solutions described in \cite{BruckmannNogvanB2003constituent,cork2018symmrot}, for which the only known examples of skyrmions exhibiting similar symmetries are actually periodic skyrmions, or \textit{Skyrme chains} \cite{HarlandWard2008chains}.
\section*{Acknowledgements}
Many thanks to Derek Harland for encouraging me to study this topic, and for the numerous discussions and correspondences we have had regarding the results of this paper.
\appendix
\section{Integrals of the ultraspherical function}\label{appendix-integrals}
In this section we shall derive formulae for the integrals
\begin{align}
    I_r=\int_{-\frac{1}{2}}^{\frac{1}{2}}\left(\phi_+^{(\alpha,1)}(t)\right)^rdt,
\end{align}
for $r=0,1,2,3,4$, where $\phi_+^{(\alpha,1)}(t)$ is given by the formula (\ref{phi_+}) as
\begin{align*}
    \phi_+^{(\alpha,1)}(t)&=\dfrac{1}{2}+2t\dfrac{\Gamma(\alpha+\frac{3}{2})}{\sqrt{\pi}\Gamma(\alpha+1)}\;{}_2F_1\left(\frac{1}{2},-\alpha;\frac{3}{2};4t^2\right).
\end{align*}
This function may be represented by the power series
\begin{align}\label{phi_+-series_2}
\phi_+^{(\alpha,1)}(t)=\dfrac{1}{2}+\mathcal{K}\sum_{k=0}^{\infty}\begin{pmatrix}\alpha\\k\end{pmatrix}\dfrac{(-1)^k}{2k+1}(2t)^{2k+1},
\end{align}
where $\mathcal{K}$ is a constant dependent only on $\alpha$, namely
\begin{align}\label{K-param}\mathcal{K}=\dfrac{\Gamma(\alpha+\frac{3}{2})}{\sqrt{\pi}\Gamma(\alpha+1)},\end{align}
and we have introduced the generalised binomial coefficient
\begin{align}\label{gen-coeff}
    \begin{pmatrix}\alpha\\k\end{pmatrix}=\dfrac{1}{k!}\left(\prod_{r=0}^{k-1}(\alpha-r)\right),
\end{align}
defined for all $\alpha\in\R$. The hypergeometric function appearing in (\ref{phi_+}) is defined in general as
\begin{align}\label{hypergeometric-function}
_pF_q(a_1,\dots,a_p;b_1,\dots,b_q;x)=\dfrac{\Gamma(b_1)\cdots\Gamma(b_q)}{\Gamma(a_1)\cdots\Gamma(a_p)}\sum_{n=0}^\infty{\dfrac{\Gamma(a_1+n)\cdots\Gamma(a_p+n)}{\Gamma(b_1+n)\cdots\Gamma(b_q+n)}}\dfrac{x^n}{n!},  
\end{align}
for non-negative integers $p\leq q+1$, and real coefficients $a_r,b_r$.

To calculate the integrals $I_r$, it is helpful to introduce the function $\eta_\alpha(t)=\phi_+^{(\alpha,1)}(t)-\frac{1}{2}$. From (\ref{phi_+}), we have that $\eta_\alpha$ is a monotone, odd function, which satisfies $\eta(\pm\frac{1}{2})=\pm\frac{1}{2}$. Introducing the notation
$$j_r(\alpha)=\int_{-\frac{1}{2}}^{\frac{1}{2}}\eta_\alpha(t)^rdt,$$
we can use the symmetry of the interval $[-1/2,1/2]$, and the fact that $\eta_\alpha$ is an odd function, to see that $j_{2n+1}=0$, and also
\begin{align}
    I_0(\alpha)&=1,\quad I_1(\alpha)=\dfrac{1}{2},\\
    I_2(\alpha)&=\dfrac{1}{4}+j_2(\alpha),\label{I_2-int}\\
    I_3(\alpha)&=\dfrac{3}{2}I_2(\alpha)-\dfrac{1}{4},\label{I_3-int}\\
    I_4(\alpha)&=\dfrac{3}{2}I_2(\alpha)-\dfrac{5}{16}+j_4(\alpha)\label{I_4-int}.
\end{align}
It thus remains to calculate $j_2$ and $j_4$.
\subsection{Evaluating \texorpdfstring{$j_2$}{j2} and \texorpdfstring{$I_2$}{I2}}
From the power series expansion (\ref{phi_+-series}), we have that
\begin{align*}
    \eta_\alpha(t)^2=\mathcal{K}^2\sum_{k=0}^\infty\begin{pmatrix}
    \alpha\\ k\end{pmatrix}\dfrac{(-1)^k}{2k+1}\sum_{j=0}^\infty\begin{pmatrix}
    \alpha\\ j\end{pmatrix}\dfrac{(-1)^j}{2j+1}(2t)^{2(k+j+1)}.
\end{align*}
Thus,
\begin{align}
    j_2(\alpha)=\mathcal{K}^2\sum_{k=0}^\infty\begin{pmatrix}
    \alpha\\ k\end{pmatrix}\dfrac{(-1)^k}{2k+1}\sum_{j=0}^\infty\begin{pmatrix}
    \alpha\\ j\end{pmatrix}\dfrac{(-1)^j}{(2j+1)(2k+2j+3)}.\label{j2-start}
\end{align}
To evaluate the sums in (\ref{j2-start}), we are going to need the following important formula:
\begin{align}\label{beta_int-2}
    \dfrac{\Gamma(z)\Gamma(w)}{\Gamma(z+w)}=2\int_{0}^{\frac{\pi}{2}}\sin^{2z-1}(x)\cos^{2w-1}(x)dx,\quad\text{for }\Re(z),\Re(w)>0.
\end{align}
With this, and the general binomial theorem
\begin{align}\label{binomial}
    (1+t)^\alpha=\sum_{r=0}^\infty\begin{pmatrix}
    \alpha\\ r\end{pmatrix}t^r,
\end{align}
we may prove the following lemma.
\begin{lemma}\label{lemma-sums}
The following sums hold for all $\alpha>-\frac{1}{2}$, $m>-1$, and $x>1$:
\begin{enumerate}
    \item $\sum_{r=0}^\infty\begin{pmatrix}
    \alpha\\ r\end{pmatrix}\dfrac{(-1)^r}{2r+1+m}=\dfrac{1}{2}\dfrac{\Gamma(\alpha+1)\Gamma(\frac{m}{2}+\frac{1}{2})}{\Gamma(\alpha+\frac{3}{2}+\frac{m}{2})}$;
    \item $\sum_{r=0}^\infty\begin{pmatrix}
    \alpha\\ r\end{pmatrix}\dfrac{(-1)^r}{r+1}=\dfrac{1}{\alpha+1}$;
    \item $\sum_{r=0}^\infty\begin{pmatrix}
    \alpha\\ r\end{pmatrix}\dfrac{(-1)^r}{(r+1)(2r+1)}=\dfrac{1}{\mathcal{K}}-\dfrac{1}{\alpha+1}$;
    \item $\sum_{r=0}^\infty\begin{pmatrix}
    \alpha\\ r\end{pmatrix}\dfrac{(-1)^r\Gamma(\alpha+1)\Gamma(r+x)}{(r+1)(r+x-1)\Gamma(\alpha+r+x+1)}=\dfrac{\Gamma(x-1)}{(2-x)(\alpha+1)}\left(\dfrac{2\Gamma(2\alpha+2)}{\Gamma(2\alpha+x+1)}-\dfrac{\Gamma(\alpha+1)}{\Gamma(\alpha+x)}\right)$.
\end{enumerate}
\end{lemma}
\textit{Proof}.
\begin{enumerate}
    \item Using the fact that by (\ref{binomial}), the sum may be written as an integral, then the substitution $t=\sin x$, and the formula (\ref{beta_int-2}), we obtain the result:
\begin{align*}
\sum_{r=0}^\infty\begin{pmatrix}
    \alpha\\ r\end{pmatrix}\dfrac{(-1)^r}{2r+1+m}&=\int_0^1(1-t^2)^\alpha t^mdt=\int_0^{\frac{\pi}{2}}\cos^{2\alpha+1}(x)\sin^m(x)dx\\
    &=\int_0^{\frac{\pi}{2}}\cos^{2(\alpha+1)-1}(x)\sin^{2(\frac{m+1}{2})-1}(x)dx=\dfrac{1}{2}\dfrac{\Gamma(\alpha+1)\Gamma(\frac{m}{2}+\frac{1}{2})}{\Gamma(\alpha+\frac{3}{2}+\frac{m}{2})}.
\end{align*}
\item First note that by (\ref{gen-coeff}), we have
$$\begin{pmatrix}\alpha+1\\r+1\end{pmatrix}=\dfrac{\alpha+1}{r+1}\begin{pmatrix}\alpha\\r\end{pmatrix}.$$
Hence, by (\ref{binomial}), we have
\begin{align*}
    0=(1-1)^{\alpha+1}&=\sum_{r=0}^\infty\begin{pmatrix}
    \alpha+1\\ r\end{pmatrix}(-1)^r=1+\sum_{r=1}^\infty\begin{pmatrix}
    \alpha+1\\ r\end{pmatrix}(-1)^r\\
    &=1-\sum_{r=0}^\infty\begin{pmatrix}
    \alpha+1\\ r+1\end{pmatrix}(-1)^r=1-(\alpha+1)\sum_{r=0}^\infty\begin{pmatrix}
    \alpha\\ r\end{pmatrix}\dfrac{(-1)^r}{r+1}.
\end{align*}
Thus
\begin{align*}
    \sum_{r=0}^\infty\begin{pmatrix}
    \alpha\\ r\end{pmatrix}\dfrac{(-1)^r}{r+1}=\dfrac{1}{\alpha+1}.
\end{align*}
\item We may split the denominator into partial fractions as
\begin{align}\label{partial-fractions-1}
    \dfrac{1}{(r+1)(2r+1)}=\dfrac{2}{2r+1}-\dfrac{1}{r+1}.
\end{align}
The result hence follows by part 1 with $m=0$, part 2, and the formula (\ref{K-param}).
\item First we note the following identities: for all $a,b,c\in\R$ such that $c\notin\Z\setminus\Z^+$ and $c-a-b>0$
\begin{align}\label{special-hyp-geom-formula}
    {}_2F_1(a,b;c;1)=\dfrac{\Gamma(c)\Gamma(c-a-b)}{\Gamma(c-a)\Gamma(c-b)}.
\end{align}
(\ref{special-hyp-geom-formula}) is also known as \textit{Gauss' hypergeometric identity} \cite{AbramowitzStegun1964}. Additionally, by (\ref{gen-coeff}), we have
\begin{align}\label{Gamma-binomial-convert}
    (-1)^r\begin{pmatrix}\alpha\\r\end{pmatrix}=\dfrac{1}{r!}\prod_{k=0}^{r-1}(k-\alpha)=\dfrac{\Gamma(r-\alpha)}{r!\Gamma(-\alpha)}.
\end{align}
Now, noting that
$$\dfrac{1}{(r+1)(r+x-1)}=\dfrac{1}{2-x}\left(\dfrac{2}{2r+2x-2}-\dfrac{1}{r+1}\right),$$
we have
\begin{align}\label{sum_4-split}
    \sum_{r=0}^\infty\begin{pmatrix}
    \alpha\\ r\end{pmatrix}\dfrac{(-1)^r}{(r+1)(r+x-1)}\dfrac{\Gamma(\alpha+1)\Gamma(r+x)}{\Gamma(\alpha+r+x+1)}=\dfrac{\Gamma(\alpha+1)}{2-x}\left(s_1(\alpha)-s_2(\alpha)\right),
\end{align}
where
\begin{align}
    s_1(\alpha)&=2\sum_{r=0}^\infty\begin{pmatrix}
    \alpha\\ r\end{pmatrix}\dfrac{(-1)^r}{2r+2x-2}\dfrac{\Gamma(r+x)}{\Gamma(\alpha+r+x+1)},\\
    s_2(\alpha)&=\sum_{r=0}^\infty\begin{pmatrix}
    \alpha\\ r\end{pmatrix}\dfrac{(-1)^r}{r+1}\dfrac{\Gamma(r+x)}{\Gamma(\alpha+r+x+1)}
\end{align}
Hence using (\ref{Gamma-binomial-convert}), and the formulae (\ref{hypergeometric-function}) and (\ref{special-hyp-geom-formula}), we have
\begin{align}
    s_1(\alpha)&=2\sum_{r=0}^\infty\dfrac{\Gamma(r-\alpha)\Gamma(r+x)}{(2r+2x-2)\Gamma(-\alpha)\Gamma(\alpha+r+x+1)}\dfrac{1}{r!}=\sum_{r=0}^\infty\dfrac{\Gamma(r-\alpha)\Gamma(r+x-1)}{\Gamma(-\alpha)\Gamma(\alpha+r+x+1)}\dfrac{1}{r!}\notag\\
    &=\dfrac{\Gamma(x-1)}{\Gamma(\alpha+x+1)}\:{}_2F_1\left(-\alpha,x-1;\alpha+x+1;1\right)=\dfrac{\Gamma(x-1)\Gamma(2\alpha+2)}{\Gamma(\alpha+2)\Gamma(2\alpha+x+1)}.\label{s-1}
\end{align}
Similarly, we have
\begin{align*}
s_2(\alpha)&=\sum_{r=0}^\infty\begin{pmatrix}
    \alpha+1\\ r+1\end{pmatrix}\dfrac{(-1)^r}{\alpha+1}\dfrac{\Gamma(r+x)}{\Gamma(\alpha+r+x+1)}\\
    &=\dfrac{\Gamma(x-1)}{(\alpha+1)\Gamma(\alpha+x)}+\sum_{r=-1}^\infty\begin{pmatrix}
    \alpha+1\\ r+1\end{pmatrix}\dfrac{(-1)^r}{\alpha+1}\dfrac{\Gamma(r+x)}{\Gamma(\alpha+r+x+1)}\\
    &=\dfrac{\Gamma(x-1)}{(\alpha+1)\Gamma(\alpha+x)}-\sum_{r=0}^\infty\begin{pmatrix}
    \alpha+1\\ r\end{pmatrix}\dfrac{(-1)^r}{\alpha+1}\dfrac{\Gamma(r+x-1)}{\Gamma(\alpha+r+x)}\\
    &=\dfrac{\Gamma(x-1)}{(\alpha+1)\Gamma(\alpha+x)}-\sum_{r=0}^\infty\dfrac{1}{\alpha+1}\dfrac{\Gamma(r-\alpha-1)\Gamma(r+x-1)}{\Gamma(-\alpha-1)\Gamma(\alpha+r+x)}\dfrac{1}{r!}\\
    &=\dfrac{\Gamma(x-1)}{(\alpha+1)\Gamma(\alpha+x)}-\dfrac{\Gamma(x-1)}{(\alpha+1)\Gamma(\alpha+x)}\:{}_2F_1\left(-\alpha-1,x-1;\alpha+x;1\right)\\
    &=\dfrac{\Gamma(x-1)}{(\alpha+1)\Gamma(\alpha+x)}-\dfrac{\Gamma(x-1)\Gamma(2\alpha+2)}{\Gamma(2\alpha+x+1)\Gamma(\alpha+2)}.
\end{align*}
Combining this with (\ref{s-1}) and (\ref{sum_4-split}) we obtain the result.
\end{enumerate}
\hfill$\square$\\

\noindent Now, to proceed in evaluating (\ref{j2-start}), we note that
$$\dfrac{1}{(2j+1)(2k+2j+3)}=\dfrac{1}{2(k+1)}\left(\dfrac{1}{2j+1}-\dfrac{1}{2j+2k+3}\right).$$
Thus by lemma \ref{lemma-sums} part 1, with $m=0$ and $m=2k+2$ respectively, we obtain
\begin{align*}
    j_2(\alpha)=\dfrac{\mathcal{K}^2}{4}\sum_{k=0}^\infty\begin{pmatrix}
    \alpha\\ k\end{pmatrix}\dfrac{(-1)^k}{(2k+1)(k+1)}\left(\dfrac{1}{\mathcal{K}}-\dfrac{\Gamma(\alpha+1)\Gamma(k+\frac{3}{2})}{\Gamma(\alpha+k+\frac{5}{2})}\right).
\end{align*}
This sum may also be split up accordingly, and evaluated using lemma \ref{lemma-sums} parts 3 and 4 (with $x=3/2$) to obtain
\begin{align}
    j_2(\alpha)&=\dfrac{\mathcal{K}^2}{4}\left(\dfrac{1}{\mathcal{K}}\left(\dfrac{1}{\mathcal{K}}-\dfrac{1}{\alpha+1}\right)-\dfrac{\sqrt{\pi}}{\alpha+1}\left(\dfrac{2\Gamma(2\alpha+2)}{\Gamma(2\alpha+\frac{5}{2})}-\dfrac{\Gamma(\alpha+1)}{\Gamma(\alpha+\frac{3}{2})}\right)\right)\notag\\
    &=\dfrac{1}{4}\left(1-\dfrac{2\Gamma(2\alpha+2)\Gamma(\alpha+\frac{3}{2})^2}{\Gamma(2\alpha+\frac{5}{2})\Gamma(\alpha+1)\Gamma(\alpha+2)\sqrt{\pi}}\right),\label{j2}
\end{align}
where we have additionally made use of the formula (\ref{K-param}) for $\mathcal{K}$. Combining (\ref{K-param}), (\ref{I_2-int}) and (\ref{j2}), we hence have that
\begin{align}
    I_2(\alpha)=\dfrac{1}{2}\left(1-\dfrac{\Gamma(2\alpha+2)\Gamma(\alpha+\frac{3}{2})^2}{\Gamma(2\alpha+\frac{5}{2})\Gamma(\alpha+1)\Gamma(\alpha+2)\sqrt{\pi}}\right).\label{I_2-appendix}
\end{align}
\subsection{Partial formulae for \texorpdfstring{$j_4$}{j4} and \texorpdfstring{$I_4$}{I4}}
From the power series expansion (\ref{phi_+-series}), we have that
\begin{align*}
    \eta_\alpha(t)^4=\mathcal{K}^4\sum_{j=0}^\infty\begin{pmatrix}
    \alpha\\ j\end{pmatrix}\dfrac{(-1)^j}{2j+1}\sum_{k=0}^\infty\begin{pmatrix}
    \alpha\\ k\end{pmatrix}\dfrac{(-1)^k}{2k+1}\sum_{l=0}^\infty\begin{pmatrix}
    \alpha\\ l\end{pmatrix}\dfrac{(-1)^l}{2l+1}\sum_{m=0}^\infty\begin{pmatrix}
    \alpha\\ m\end{pmatrix}\dfrac{(-1)^m}{2m+1}(2t)^{2(j+k+l+m+2)}.
\end{align*}
Thus,
\begin{align}
    j_4(\alpha)=\mathcal{K}^4\sum_{j=0}^\infty r_j(\alpha)\sum_{k=0}^\infty r_k(\alpha)\sum_{l=0}^\infty r_l(\alpha)\sum_{m=0}^\infty\begin{pmatrix}
    \alpha\\ m\end{pmatrix}\dfrac{(-1)^m}{(2m+1)(2j+2k+2l+2m+5)},\label{j4-start}
\end{align}
where we have introduced the notation
$$r_j(\alpha)=\begin{pmatrix}
    \alpha\\ j\end{pmatrix}\dfrac{(-1)^j}{2j+1}$$
to simplify expressions. The first step in evaluating (\ref{j4-start}) is to note that
$$\dfrac{1}{(2m+1)(2j+2k+2l+2m+5)}=\dfrac{1}{2(j+k+l+2)}\left(\dfrac{1}{2m+1}-\dfrac{1}{2j+2k+2l+2m+5}\right).$$
Hence, by lemma \ref{lemma-sums} part 1 with $m=0$ and $m=2j+2k+2l+4$ respectively, we have
\begin{align}
    j_4(\alpha)&=\dfrac{\mathcal{K}^4}{2}\sum_{j,k=0}^\infty r_j(\alpha)r_k(\alpha)\sum_{l=0}^\infty\dfrac{r_l(\alpha)}{2j+2k+2l+4}\left(\dfrac{1}{\mathcal{K}}-\dfrac{\Gamma(\alpha+1)\Gamma(j+k+l+\frac{5}{2})}{\Gamma(\alpha+j+k+l+\frac{7}{2})}\right)\notag\\
    &=\dfrac{\mathcal{K}^4}{2}\sum_{j,k=0}^\infty r_j(\alpha)r_k(\alpha)\left(\dfrac{\sigma_1(\alpha,j+k)}{\mathcal{K}}-\sigma_2(\alpha,j+k)\right),\label{j4-mid}
\end{align}
where we have defined the sums
\begin{align}
    \sigma_1(\alpha,m)&=\sum_{r=0}^\infty\begin{pmatrix}\alpha\\r\end{pmatrix}\dfrac{(-1)^r}{(2r+1)(2m+2r+4)},\label{s_1}\\
    \sigma_2(\alpha,m)&=\sum_{r=0}^\infty\begin{pmatrix}\alpha\\r\end{pmatrix}\dfrac{(-1)^r}{(2r+1)(2m+2r+4)}\dfrac{\Gamma(\alpha+1)\Gamma(m+r+\frac{5}{2})}{\Gamma(\alpha+m+r+\frac{7}{2})}.\label{s_2}
\end{align}
The sum (\ref{s_1}) can be evaluated straightforwardly by using lemma \ref{lemma-sums} part 1. Indeed, noting that
\begin{align}\label{partial-fraction-for-sigmas}
    \dfrac{1}{(2r+1)(2m+2r+4)}=\dfrac{1}{2m+3}\left(\dfrac{1}{2r+1}-\dfrac{1}{2m+2r+4}\right),
\end{align}
we have that (\ref{s_1}) reduces to
\begin{align}
    \sigma_1(\alpha,m)=\dfrac{1}{2m+3}\left(\dfrac{1}{2\mathcal{K}}-\dfrac{1}{2}\dfrac{\Gamma(\alpha+1)\Gamma(m+2)}{\Gamma(\alpha+m+3)}\right).\label{s1}
\end{align}
For the sum (\ref{s_2}), we need some more formulae.
\begin{lemma}\label{lemma-sum-of-gammas-hyper}
The following holds for all $\alpha>-\frac{1}{2}$, and $x,y>0$.
$$\sum_{r=0}^\infty\begin{pmatrix}\alpha\\r\end{pmatrix}\dfrac{(-1)^r}{2r+2y}\dfrac{\Gamma(\alpha+1)\Gamma(r+x)}{\Gamma(\alpha+r+x+1)}=\dfrac{y}{2}\dfrac{\Gamma(\alpha+1)\Gamma(x)}{\Gamma(\alpha+x+1)}\:{}_3F_2\left(-\alpha,x,y;\alpha+x+1,y+1;1\right).$$
\end{lemma}
\textit{Proof}. This all follows straightforwardly by the formulae (\ref{hypergeometric-function}) and (\ref{Gamma-binomial-convert}):
\begin{align*}
\sum_{r=0}^\infty\begin{pmatrix}\alpha\\r\end{pmatrix}\dfrac{(-1)^r}{2r+2y}\dfrac{\Gamma(\alpha+1)\Gamma(r+x)}{\Gamma(\alpha+r+x+1)}&=\dfrac{\Gamma(\alpha+1)}{2}\sum_{r=0}^\infty\dfrac{\Gamma(r-\alpha)\Gamma(x+r)\Gamma(r+y)}{\Gamma(r+y+1)\Gamma(\alpha+x+1+r)\Gamma(-\alpha)r!}\\
&=\dfrac{\Gamma(\alpha+1)\Gamma(x)\Gamma(y)}{2\Gamma(\alpha+x+1)\Gamma(y+1)}\:{}_3F_2\left(-\alpha,x,y;\alpha+x+1,y+1;1\right)\\
&=\dfrac{y}{2}\dfrac{\Gamma(\alpha+1)\Gamma(x)}{\Gamma(\alpha+x+1)}\:{}_3F_2\left(-\alpha,x,y;\alpha+x+1,y+1;1\right).
\end{align*}
\hfill$\square$\\

\noindent Therefore, utilising the formula (\ref{partial-fraction-for-sigmas}), and lemma \ref{lemma-sum-of-gammas-hyper} with $x=m+5/2$ and both $y=m+2$ and $y=\frac{1}{2}$, the sum (\ref{s_2}) reduces to
\begin{align}
    \sigma_2(\alpha,m)=\dfrac{\Gamma(\alpha+1)\Gamma(m+\frac{5}{2})}{(2m+3)\Gamma(\alpha+m+\frac{7}{2})}&\left(\dfrac{1}{4}\:{}_3F_2\left(-\alpha,m+\frac{5}{2},\frac{1}{2};\alpha+m+\frac{7}{2},\frac{3}{2};1\right)\right.\label{s2}\\
    &\:\left.-\dfrac{m+2}{2}\:{}_3F_2\left(-\alpha,m+\frac{5}{2},m+2;\alpha+m+\frac{7}{2},m+3;1\right)\right).\notag
\end{align}
So, we now have from (\ref{j2-start}), (\ref{j4-mid}), (\ref{s1}), and (\ref{s2}), that
\begin{align}
    j_4(\alpha)=&\:\dfrac{1}{4}j_2(\alpha)+\Gamma(\alpha+1)\dfrac{\mathcal{K}^4}{8}\Sigma(\alpha),
\end{align}
where
\begin{align}
    \Sigma(\alpha)=\sum_{j,k=0}^\infty \dfrac{r_j(\alpha)r_k(\alpha)}{(2j+2k+3)}\left(\mathcal{L}_1(j+k)-\mathcal{L}_2(j+k)-\dfrac{2}{\mathcal{K}}\dfrac{\Gamma(j+k+2)}{\Gamma(\alpha+j+k+3)}\right),\label{Sigma}
\end{align}
with
\begin{align}
    \mathcal{L}_1(m)&=\dfrac{(2m+4)\Gamma(m+\frac{5}{2})}{\Gamma(\alpha+m+\frac{7}{2})}\:{}_3F_2\left(-\alpha,m+\frac{5}{2},m+2;\alpha+m+\frac{7}{2},m+3;1\right),\\
    \mathcal{L}_2(m)&=\dfrac{\Gamma(m+\frac{5}{2})}{\Gamma(\alpha+m+\frac{7}{2})}\:{}_3F_2\left(-\alpha,m+\frac{5}{2},\frac{1}{2};\alpha+m+\frac{7}{2}.\frac{3}{2};1\right)
\end{align}
This sum (\ref{Sigma}) is very hard to compute due to the occurrence of the generalised hypergeometric functions. The only part we can compute is one part of the third term in the sum, namely, by lemma \ref{lemma-sum-of-gammas-hyper}, we have
\begin{align}
    &\sum_{j=0}^\infty \dfrac{r_j(\alpha)}{(2j+2k+3)}\dfrac{\Gamma(j+k+2)}{\Gamma(\alpha+j+k+3)}=\dfrac{1}{2k+2}\sum_{j=0}^\infty \begin{pmatrix}\alpha\\j\end{pmatrix}\dfrac{\Gamma(j+k+2)}{\Gamma(\alpha+j+k+3)}\left(\dfrac{(-1)^j}{2j+1}-\dfrac{(-1)^j}{2j+2k+3}\right)\notag\\
    &\qquad=\:\dfrac{1}{8(k+1)}\dfrac{\Gamma(k+2)}{\Gamma(\alpha+k+3)}\left(\:{}_3F_2\left(-\alpha,k+2,\frac{1}{2};\alpha+k+3,\frac{3}{2};1\right)\right.\\
    &\qquad\qquad\qquad\qquad\qquad\qquad\qquad\qquad\left.-(2k+3)\:{}_3F_2\left(-\alpha,k+2,k+\frac{3}{2};\alpha+k+3,k+\frac{5}{2};1\right)\right).\notag
\end{align}
\bibliographystyle{unsrt}
\bibliography{references}

\begin{thebibliography}{10}

\bibitem{MantonSutcliffe2004}
N~S Manton and P~M Sutcliffe.
\newblock {\em Topological solitons}.
\newblock Cambridge University Press, 2004.

\bibitem{shnir2018topsolitons}
Y~M Shnir.
\newblock {\em Topological and non-topological solitons in scalar field
  theories}.
\newblock Cambridge University Press, 2018.

\bibitem{skyrme1962nucl}
T~H~R Skyrme.
\newblock A unified field theory of mesons and baryons.
\newblock {\em Nucl. Phys.}, 31:556, 1962.

\bibitem{Manton1987geometry}
N~S Manton.
\newblock Geometry of skyrmions.
\newblock {\em Commun. Math. Phys.}, 111(3):469--478, 1987.

\bibitem{AlvarezCanforaDimakisPaliathanasis2017integrability}
P~D Alvarez, F~Canfora, N~Dimakis, and A~Paliathanasis.
\newblock Integrability and chemical potential in the {(3+ 1)-dimensional}
  {Skyrme} model.
\newblock {\em Phys. Lett. B}, 773:401--407, 2017.

\bibitem{esteban1986SkyrmeVarSol}
M~J Esteban.
\newblock A direct variational approach to {Skyrme's} model for meson fields.
\newblock {\em Commun. Math. Phys.}, 105(4):571--591, 1986.

\bibitem{esteban2004existence}
M~J Esteban.
\newblock Existence of {3-D} skyrmions.
\newblock {\em Commun. Math. Phys.}, 251(1):209--210, 2004.

\bibitem{battyesutcliffe1997symmetricskyme}
R~A Battye and P~M Sutcliffe.
\newblock Symmetric skyrmions.
\newblock {\em Phys. Rev. Lett.}, 79(3):363, 1997.

\bibitem{battye2001solitonic}
R~A Battye and P~M Sutcliffe.
\newblock Solitonic fullerene structures in light atomic nuclei.
\newblock {\em Phys. Rev. Lett.}, 86(18):3989, 2001.

\bibitem{battye2002skyrmions}
R~A Battye and Paul~M Sutcliffe.
\newblock Skyrmions, fullerenes and rational maps.
\newblock {\em Rev. Math. Phys.}, 14(01):29--85, 2002.

\bibitem{BraatenTownsendCarson1990}
E~Braaten, L~Carson, and S~Townsend.
\newblock Novel structure of static multisoliton solutions in the {Skyrme}
  model.
\newblock {\em Phys. Lett. B}, 235(1-2):147--152, 1990.

\bibitem{FeistLauManton2013skyrmions}
D~T~J Feist, P~H~C Lau, and N~S Manton.
\newblock Skyrmions up to baryon number {108}.
\newblock {\em Phys. Rev. D}, 87(8):085034, 2013.

\bibitem{AtiyahManton1989}
M~F Atiyah and N~S Manton.
\newblock Skyrmions from instantons.
\newblock {\em Phys. Lett. B}, 222(3):438--442, 1989.

\bibitem{LeeseManton1994stable}
R~A Leese and N~S Manton.
\newblock Stable instanton-generated {Skyrme} fields with baryon numbers three
  and four.
\newblock {\em Nucl. Phys. A}, 572(3-4):575--599, 1994.

\bibitem{MantonSutcliffe1995skyrme}
N~S Manton and P~M Sutcliffe.
\newblock {Skyrme} crystal from a twisted instanton on a four-torus.
\newblock {\em Phys. Lett. B}, 342(1-4):196--200, 1995.

\bibitem{SingerSutcliffe1999}
M~A Singer and P~M Sutcliffe.
\newblock Symmetric instantons and {Skyrme} fields.
\newblock {\em Nonlinearity}, 12(4):987, 1999.

\bibitem{sutcliffe2004Buckyball}
P~M Sutcliffe.
\newblock Instantons and the buckyball.
\newblock In {\em Proc. R. Soc. Lon. A: Mathematical, Physical and Engineering
  Sciences}, volume 460, pages 2903--2912. The Royal Society, 2004.

\bibitem{sutcliffe2010skyrmions}
P~Sutcliffe.
\newblock Skyrmions, instantons and holography.
\newblock {\em J. H. E. P.}, 2010(8):19, 2010.

\bibitem{SakaiSugimoto2005}
T~Sakai and S~Sugimoto.
\newblock Low energy hadron physics in holographic {QCD}.
\newblock {\em Prog. Th. Phys.}, 113(4):843--882, 2005.

\bibitem{sutcliffe2011skyrmionsTruncated}
P~Sutcliffe.
\newblock Skyrmions in a truncated {BPS} theory.
\newblock {\em J. H. E. P.}, 2011(4):45, 2011.

\bibitem{NayaSutcliffe2018skyrmions}
C~Naya and P~Sutcliffe.
\newblock Skyrmions in models with pions and rho mesons.
\newblock {\em J. H. E. P.}, 2018(5):174, 2018.

\bibitem{CharbonneauHurtubise2008Rat.map}
B~Charbonneau and J~Hurtubise.
\newblock Calorons, {Nahm’s} equations on {$S^1$} and bundles over
  {$\mathbb{P}^{1}\times\mathbb{P}^1$}.
\newblock {\em Commun. Math. Phys.}, 280(2):315--349, 2008.

\bibitem{GarlandMurray1988}
H~Garland and M~K Murray.
\newblock {Kac-Moody} monopoles and periodic instantons.
\newblock {\em Commun. Math. Phys.}, 120(2):335--351, 1988.

\bibitem{HekmatiMurrayVozzo2012CalCor}
P~Hekmati, M~K Murray, and R~F Vozzo.
\newblock The general caloron correspondence.
\newblock {\em J. Geom. Phys.}, 62(2):224--241, 2012.

\bibitem{KraanVanBaal1998Tdual}
T~C Kraan and P~van Baal.
\newblock Exact {T-duality} between calorons and {Taub-NUT} spaces.
\newblock {\em Phys. Lett. B}, 428(3):268--276, 1998.

\bibitem{kraanVanBaal1998monopoleconsts}
T~C Kraan and P~van Baal.
\newblock Monopole constituents inside {$SU(n)$} calorons.
\newblock {\em Phys. Lett. B}, 435(3-4):389--395, 1998.

\bibitem{KraanVanBaal1998}
T~C Kraan and P~van Baal.
\newblock Periodic instantons with non-trivial holonomy.
\newblock {\em Nucl. Phys. B}, 533(1):627--659, 1998.

\bibitem{norbury2000periodic}
P~Norbury.
\newblock Periodic instantons and the loop group.
\newblock {\em Commun. Math. Phys.}, 212(3):557--569, 2000.

\bibitem{Harland2007}
D~G Harland.
\newblock Large scale and large period limits of symmetric calorons.
\newblock {\em J. Math. Phys.}, 48(8):082905, 2007.

\bibitem{LeeLu1998}
K~Lee and C~Lu.
\newblock {$SU(2)$} calorons and magnetic monopoles.
\newblock {\em Phys. Rev. D}, 58(2):025011, 1998.

\bibitem{Rossi1979}
P~Rossi.
\newblock Propagation functions in the field of a monopole.
\newblock {\em Nucl. Phys. B}, 149(1):170--188, 1979.

\bibitem{Ward2004}
R~S Ward.
\newblock Symmetric calorons.
\newblock {\em Phys. Lett. B}, 582(3):203--210, 2004.

\bibitem{AlkoferGreensite2007quark}
R~Alkofer and J~Greensite.
\newblock Quark confinement: the hard problem of hadron physics.
\newblock {\em J. Phys. G: Nuclear and Particle Physics}, 34(7):S3, 2007.

\bibitem{vanBaal2007review}
P~van Baal.
\newblock A review of instanton quarks and confinement.
\newblock In {\em AIP Conference Proceedings}, volume 892, pages 241--244. AIP,
  2007.

\bibitem{HarlandWard2008chains}
D~G Harland and R~S Ward.
\newblock Chains of skyrmions.
\newblock {\em J. H. E. P.}, 2008(12):093, 2008.

\bibitem{EskolaKajantie1989skycal}
K~J Eskola and K~Kajantie.
\newblock Thermal skyrmion-like configuration.
\newblock {\em Zeitschrift f{\"u}r Physik C Particles and Fields},
  44(2):347--348, 1989.

\bibitem{NowakZahed1989skycal}
M~A Nowak and I~Zahed.
\newblock Skyrmions from instantons at finite temperature.
\newblock {\em Phys. Lett. B}, 230(1-2):108--112, 1989.

\bibitem{MantonSamols1990skyrmions}
N~S Manton and T~M Samols.
\newblock Skyrmions on {$S^3$} and {$H^3$} from instantons.
\newblock {\em J. Phys. A: Mathematical and General}, 23(16):3749, 1990.

\bibitem{ArthurTchrakian1996GaugedSky}
K~Arthur and D~H Tchrakian.
\newblock {$SO(3)$} gauged soliton of an {$O(4)$} sigma model on
  {$\mathbb{R}^3$}.
\newblock {\em Phys. Lett. B}, 378(1-4):187--193, 1996.

\bibitem{BrihayeHartmannTchhrakian2001MonopolesGaugeSky}
Y~Brihaye, B~Hartmann, and D~H Tchrakian.
\newblock Monopoles and dyons in {$SO(3)$} gauged {Skyrme} models.
\newblock {\em J. Math. Phys.}, 42(8):3270--3281, 2001.

\bibitem{BrihayeHillZachos2004BoundingGaugeSkyMass}
Y~Brihaye, C~T Hill, and C~K Zachos.
\newblock Bounding gauged skyrmion masses.
\newblock {\em Phys. Rev. D}, 70(11):111502, 2004.

\bibitem{PietteTchrakian2000static}
B~M A~G Piette and D~H Tchrakian.
\newblock Static solutions in the {$U(1)$} gauged {Skyrme} model.
\newblock {\em Phys. Rev. D}, 62(2):025020, 2000.

\bibitem{HoughtonMantonSutcliffe1998rational}
C~J Houghton, N~S Manton, and P~M Sutcliffe.
\newblock Rational maps, monopoles and skyrmions.
\newblock {\em Nucl. Phys. B}, 510(3):507--537, 1998.

\bibitem{GrigorievSutcliffeTchrakian2002skyrmedmon}
D~Y Grigoriev, P~M Sutcliffe, and D~H Tchrakian.
\newblock Skyrmed monopoles.
\newblock {\em Phys. Lett. B}, 540(1-2):146--152, 2002.

\bibitem{PaturyanTchrakian2004monopoleSky}
V~Paturyan and D~H Tchrakian.
\newblock Monopole--antimonopole solutions of the skyrmed {$SU(2)$}
  {Yang--Mills--Higgs} model.
\newblock {\em J. Math. Phys.}, 45(1):302--309, 2004.

\bibitem{faddeev1976some}
L~D Faddeev.
\newblock Some comments on the many-dimensional solitons.
\newblock {\em Lett. Math. Phys.}, 1(4):289--293, 1976.

\bibitem{Uhlenbeck1982}
K~K Uhlenbeck.
\newblock Removable singularities in {Yang-Mills} fields.
\newblock {\em Commun. Math. Phys.}, 83(1):11--29, 1982.

\bibitem{atiyahhitchinsinger1978self}
M~F Atiyah, N~Js Hitchin, and I~M Singer.
\newblock Self-duality in four-dimensional riemannian geometry.
\newblock {\em Proc. R. Soc. Lond. A}, 362(1711):425--461, 1978.

\bibitem{AbramowitzStegun1964}
M~Abramowitz and I~A Stegun.
\newblock {\em Handbook of mathematical functions: with formulas, graphs, and
  mathematical tables}, volume~55.
\newblock Courier Corporation, 1964.

\bibitem{Nyethesis}
T~M~W Nye.
\newblock {\em The geometry of calorons}.
\newblock PhD thesis, The University of Edinburgh, 2001.

\bibitem{donaldson1984instantons}
S~K Donaldson.
\newblock Instantons and geometric invariant theory.
\newblock {\em Commun. Math. Phys.}, 93(4):453--460, 1984.

\bibitem{HarringtonShepard1978periodic}
B~J Harrington and H~K Shepard.
\newblock Periodic euclidean solutions and the finite-temperature {Yang-Mills}
  gas.
\newblock {\em Phys. Rev. D}, 17(8):2122, 1978.

\bibitem{PrasadSommerfield1975}
M~K Prasad and C~M Sommerfield.
\newblock Exact classical solution for the{ 't Hooft} monopole and the
  {Julia-Zee} dyon.
\newblock {\em Phys. Rev. Lett.}, 35(12):760, 1975.

\bibitem{cork2018symmrot}
J~Cork.
\newblock Symmetric calorons and the rotation map.
\newblock {\em J. Math. Phys.}, 59(6):062902, 2018.

\bibitem{FloodSpeight2018chern}
S~Flood and J~M Speight.
\newblock {Chern-Simons} deformation of vortices on compact domains.
\newblock {\em J. Geom. Phys.}, 133:153--167, 2018.

\bibitem{KatoNakamulaTakesue2018magnetically}
T~Kato, A~Nakamula, and K~Takesue.
\newblock Magnetically charged calorons with non-trivial holonomy.
\newblock {\em J. H. E. P.}, 2018(6):24, 2018.

\bibitem{BruckmannNogvanB2003constituent}
F~Bruckmann, D~N{\'o}gr{\'a}di, and P~van Baal.
\newblock Constituent monopoles through the eyes of fermion zero-modes.
\newblock {\em Nucl. Phys. B}, 666(1):197--229, 2003.

\end{thebibliography}
\end{document}